\documentclass[twocolumn,prb,showpacs,preprintnumbers,superscriptaddress,amsmath,amssymb,citeautoscript]{revtex4-1}
\usepackage{graphicx}
\usepackage{dcolumn}
\usepackage{color}
\usepackage{bm}
\usepackage[hidelinks]{hyperref}
\hypersetup{
    colorlinks,
    citecolor=blue,
    filecolor=blue,
    linkcolor=blue,
    urlcolor=blue
}

\newcommand{\im}{\textrm{i}}

\DeclareMathOperator{\sgn}{sgn}

\DeclareMathOperator{\e}{e}
\DeclareMathOperator{\C}{\mathcal{C}}

\DeclareMathOperator{\CL}{\mathcal{L}}
\DeclareMathOperator{\CuL}{\underline{\mathcal{L}}}

\begin{document}

\title{Majorana bound states in open quasi-1D and 2D systems with transverse Rashba coupling}

\author{N. Sedlmayr}
\email{nicholas.sedlmayr@cea.fr}
\affiliation{Institut de Physique Th\'eorique, CEA/Saclay,
Orme des Merisiers, 91190 Gif-sur-Yvette Cedex, France}
\author{J.M. Aguiar-Hualde}
\affiliation{Institut de Physique Th\'eorique, CEA/Saclay,
Orme des Merisiers, 91190 Gif-sur-Yvette Cedex, France}
\affiliation{Laboratoire d'Etude des Microstructures, ONERA-CNRS, BP 72, F-92322, Ch\^atillon, France}
\author{C. Bena}
\affiliation{Institut de Physique Th\'eorique, CEA/Saclay,
Orme des Merisiers, 91190 Gif-sur-Yvette Cedex, France}
\affiliation{Laboratoire de Physique des Solides, UMR 8502, B\^at. 510, 91405 Orsay Cedex, France}

\date{\today}

\begin{abstract}
We study the formation of Majorana states in quasi-1D and 2D square lattices with open boundary conditions, with general anisotropic Rashba coupling, in the presence of an applied Zeeman field and in the proximity of a superconductor. For systems in which the length of the system is very large (quasi-1D)  we calculate analytically the exact topological invariant, and we find a rich corresponding phase diagram which is strongly dependent on the width of the system. We compare our results with previous results based on a few-band approximation. We also investigate numerically open 2D systems of finite length in both directions. We use the recently introduced generalized Majorana polarization, which can locally evaluate  the Majorana character of a given state.  We find that the formation of Majoranas depends strongly on the geometry of the system: for a very elongated wire the finite-size numerical phase diagram reproduces the analytical phase diagram for infinite systems, while if  the length and the width are comparable no Majorana states can form, however, one can show the formation of ``quasi-Majorana'' states that have a local Majorana character, but no global Majorana symmetry.
\end{abstract}


\maketitle

\section{Introduction}\label{sec_int}

Many theoretical works have predicted the existence of Majorana fermions in various solid state systems\cite{Qi2011}, for example in one-dimensional semiconducting wires with either strong spin-orbit coupling in the presence of an uniform Zeeman magnetic field\cite{Kitaev2001,Fu2008,Fu2009,Sato2009,Lutchyn2010,Oreg2010}, or without spin orbit but with an inhomogenous Zeeman field\cite{Yazdani1997,Choy2011,Kjaergaard2012,Martin2012,Tewari2012,Klinovaja2012a,Klinovaja2013,Nadj-Perge2013,Sau2013,Pientka2013,Ben-Shach2015}, as well as in various two-dimensional systems and arrays\cite{Potter2010,Mizushima2013,Wang2014,Poyhonen2014,Seroussi2014,Wakatsuki2014,Deng2014,San-Jose2014,Mohanta2014,Sedlmayr2015}. Though there have already been several promising experiments \cite{Mourik2012,Deng2012,Das2012,Lee2014,Nadj-Perge2014} their interpretation is controversial and no clear-cut detection of Majoranas has so far been attained. An important aspect to understand in order to connect with the experimental measurements is the effect of the geometry of a system in the formation of Majorana states. Most previous works dealing with the formation of Majoranas have focused on either one-dimensional finite-size systems or on two-dimensional nanoribbons with periodic boundary conditions, with a few exceptions\cite{Potter2010}, some of them focusing on multi-channel one-dimensional wires \cite{Potter2011,Law2011,Lutchyn2011,Lutchyn2011a,Stanescu2011,Lim2012,Stanescu2013,Wang2015a,Tewari2012,Pekerten2015}. 

Here we are interested in the effect of the geometry of a system on the conditions to form Majorana states in quasi-1D and 2D systems. We focus on two main questions: what are the conditions for the formation of Majoranas 1) in quasi-1D infinite systems with both longitudinal and transverse Rashba coupling, and 2) in systems with open boundary conditions in both directions, also in the presence of both longitudinal and transverse Rashba.

To answer the first question, previous works have focused on using multi-band approximations\cite{Potter2011,Law2011,Stanescu2011,Lim2012,Stanescu2013,Wang2015a}, or on perturbing a simpler underlying model\cite{Tewari2012}, both approaches limiting though the applicability of the results to a small region of parameter space. 
Here we use a different technique which allows us to go beyond these approximation and calculate exactly the topological invariant for a system with arbitrary longitudinal and transversal hopping and spin-orbit coupling, using a fully analytical technique similar to the one presented in Ref.~\onlinecite{Dutreix2014b}. This calculation allows us to determine the topological phase diagram for such systems which is very rich and depends strongly on the width of the wire. Moreover, we note that, in contrast to the single channel model the strength of the spin-orbit coupling now enters explicitly into the form of the invariant and affects the form of the phase diagram. We compare our exact results with results obtained using an approximation of our system to a multi-band one-dimensional system, which is however valid only for a small region in the parameter space. We can show that in this region of the parameter space the exact topological phase diagram is recovered accurately by the multi-band approximation, however our theory is much more general and can describe the topological phase diagram for the entire parameter space.

The second question that we address, the condition for the formation of Majoranas in fully open systems of arbitrary size, has, to our knowledge, not been previously addressed quantitatively for realistic models. Previous studies have focused on purely one-dimensional systems, two-dimensional systems with periodic boundary conditions  (i.e. infinite nanoribbons)  \cite{Sato2009a,Sato2009b}, as well as open systems in which the aspect ratios are very large (quasi-1D) by using certain approximations, as mentioned above\cite{Potter2010,Potter2011,Law2011,Stanescu2011,Lim2012,Stanescu2013,Wang2015a,Tewari2012}.  However, the case of the fully open systems of arbitrary size  has not been previously addressed, and the analysis of such system is much more complex, as we expect that in the presence of fully open boundary conditions the Majorana states would connect and hybridize, and thus get destroyed. This phenomenon has already been seen in the quantum Hall effect.\cite{Qi2006} 
Moreover, such analysis is more relevant than the previous studies for the actual experimental situations\cite{Mourik2012} , since the wires studied in the transport experiments are neither purely one-dimensional, nor infinite ribbons, but open quasi-2D or quasi-3D systems with a finite transverse width for which transversal hopping, as well as transversal Rashba coupling are important. 

To explore this complex problem we vary the ratio between the length and the width of our system and we analyze the low-energy states numerically, using the MathQ code\footnote{See http://www.icmm.csic.es/sanjose/MathQ/MathQ.html}. 
A crucial tool in our analysis is the Majorana polarization (MP) whose general definition was recently introduced in Ref.~\onlinecite{Sedlmayr2015b}; this plays the role of a local measure of the Majorana character of a given state. Previous studies have focused only on the local density of states, which carries no direct information about the Majorana nature of a state\cite{Potter2010}. Using the MP we find that the formation of Majorana states in fully open systems is strongly affected by the geometry of the system. Thus, for asymmetric elongated systems, i.e.~the quasi-1D limit, we recover numerically the phase diagrams that we had obtained also analytically by calculating the topological invariant. When the system is deformed towards a quasi-square system, the Majorana states are gradually destroyed for larger and larger regions in the parameter space, and for a square system Majorana states cannot form, however low-energy quasi-Majorana states exist. Such states have locally an almost perfect Majorana character, however the Majorana polarization vector is not aligned but varies spatially and thus a full Majorana state cannot form. Such states have also been observed in the spinless Kitaev model \cite{Sedlmayr2015b}. The parameter range in which such quasi-Majorana states form corresponds surprisingly enough to the 2D bulk topological phase predicted for this system \cite{Sato2009a,Sato2009b,Sato2010}. Note that these states are similar to the `chiral' Majorana states \cite{Potter2010,Alicea2012}, which are also Majorana-like low-(but non-zero)-energy states arising on the internal and external boundaries of a finite size system with a hole. These `chiral' Majorana states would become exact Majoranas when a magnetic flux is threaded through the system.
The relationship between the quasi-Majorana states and the chiral Majorana states is the topic of a separate investigation.

We should note that the two approaches described above for the two questions addressed in this work are very general and can be generalized to predict the conditions for Majorana formation for systems of arbitrary size, dimension and lattice structure, as well as for any type of hopping and spin-orbit couplings. While we focus here on a rather simple tight-binding model applying to a two-dimensional square lattice, we should stress that our approach is fully generalizable to more complex tight-binding models that can map more accurately realistic systems such as InAs and InSb wires; a full numerical analysis of such more complex models is envisageable in the future.
 
This paper is organized as follows. In Sec.~\ref{sec_model} we introduce the general model. In Sec.~\ref{sec_topinv} we calculate analytically the topological invariant for a generic infinite quasi-1D systems. In Sec.~\ref{sec_2D} we present the numerical phase diagrams obtained for finite-size systems of different geometries using the Majorana polarization, and we show that in the limit of very elongated wires this coincides with the phase diagram computed analytically in Sec.~\ref{sec_topinv}. In Sec.~\ref{sec_dos} we look at the signatures of the phase transitions in experimentally measurable quantities, such as the density of states. In Sec.~\ref{sec_multiband} we explore the connection with the previously studied multi-band models and we show how we can recover the same results for specific regions in the parameter space. We conclude in Sec.~\ref{sec_conclusions}. More detailed examples are presented in the Appendices, such as the effects of the asymmetry between the longitudinal and transverse Rashba, as well as the transition between open and periodic boundary conditions.

\section{Model}\label{sec_model}

We first introduce the Hamiltonian which underlies our analysis. We use the Nambu basis:  $\Psi^\dagger_{\vec r}=\{c^\dagger_{{\vec r}\uparrow},c^\dagger_{{\vec r}\downarrow},c_{{\vec r}\downarrow},-c_{{\vec r}\uparrow}\}$, where $c_{{\vec r}\sigma}^{(\dagger)}$ annihilates (creates) a particle of spin $\sigma$ at site ${\vec r}=(i,j)$ in a square lattice. The corresponding wavefunction is $\psi^T_{{\vec r}}$: $\{u_{{\vec r}\uparrow},u_{{\vec r}\downarrow},v_{{\vec r}\downarrow},v_{{\vec r}\uparrow}\}$. We also use $\vec{\bm\tau}$ to denote the Pauli matrices in the particle-hole subspace and $\vec{\bm\sigma}$ for the Pauli matrices in the spin subspace.
We consider the two dimensional model Hamiltonian
\begin{eqnarray}\label{hamiltonian}
H=&&\sum_{{\vec r}}\bigg[\Psi^\dagger_{{\vec r}}\left(-\mu{\bm\tau}^z-\Delta{\bm\tau}^x+B{\bm\sigma}^z\right)\Psi_{{\vec r}}\nonumber
\\&&+\Psi^\dagger_{{\vec r}}\left(-t_x-\im\alpha_x{\bm\sigma}^y\right){\bm \tau}^z\Psi_{{\vec r}+\hat x}+\textrm{H.c.}\nonumber
\\&&+\Psi^\dagger_{{\vec r}}\left(-t_y+\im\alpha_y{\bm\sigma}^x\right){\bm \tau}^z\Psi_{{\vec r}+\hat y}+\textrm{H.c.}\bigg]\,,
\end{eqnarray}
$t_{x,y}$ is the nearest neighbour hopping,
$\mu$ the chemical potential, $B$ the magnetic field, $\Delta$ the proximity induced s-wave superconducting pairing, and
$\alpha_{x,y}$ are the strength of the longitudinal and transversal Rashba spin-orbit couplings.  We set $t_x=t=1$ and $\hbar=1$ throughout. Unless explicitly stated we also set $t_y=t_x=1$. Here $\hat x$ and $\hat y$ are the unit vectors for the $x$ and $y$ directions respectively, and the lattice spacing is $a=1$. We will focus exclusively on square lattices. We also define the particle hole operator $\C=\e^{\im\zeta}{\bm \sigma}^y{\bm \tau}^y\hat K$, where $\hat K$ is the complex-conjugation operator, and $\zeta$ is an arbitrary phase. The Hamiltonian anti-commutes with this operator, $\{\C,H\}=0$, and $\C^2=1$.

The size of the system to be analyzed is  $N_x\times N_y$. For the quasi-1D systems $N_x\gg N_y$. The magnetic field is taken to be along $\hat z$.
For the 2D open systems we consider here a magnetic field perpendicular to the plane and we neglect the orbital effects; the orientation of the magnetic field may affect the results but we do not focus on these effects in this manuscript, but it is the focus of a separate investigation.

\begin{figure}
\includegraphics*[width=0.95\columnwidth]{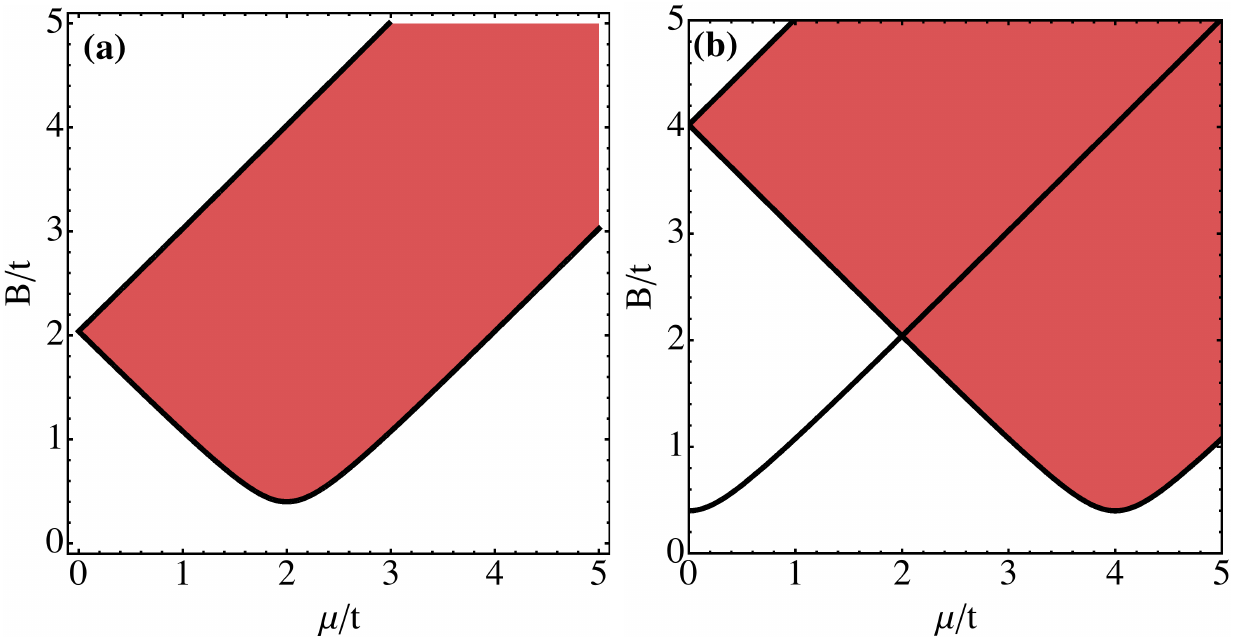}
\caption{(Color online) Bulk topological phase diagram for (a) a strictly 1D -  single channel, wire and (b) a square lattice, where $\Delta=0.4t$. White is the topologically trivial phase, light red is the bulk topologically non-trivial phase where a pair of Majorana edge states form at the boundaries. The black lines show  where the bulk gap closes.}
\label{figure1}
\end{figure}

Two limits are already well understood. If $N_y=1$ we are back to a strictly 1D wire. If both the $x$ and $y$ directions are bulk, one has the standard 2D case.\cite{Kitaev2001,Sato2009,Sato2009a,Sato2010a,Sato2010} The topological phase diagram of both of these cases is already known, see Fig.~\ref{figure1}. The 2D array is in the topological class D described by a Chern number, a $\mathbb{Z}$ invariant. The 1D case is in a higher symmetry class BDI\cite{Schnyder2008} due to the absence of $\alpha_y$ and is also described by a $\mathbb{Z}$ invariant, a winding number. The introduction of $\alpha_y$ in the quasi-1D breaks a so-called time-reversal symmetry putting it into the D class which in 1D has a $\mathbb{Z}_2$ invariant, calculated from the Pfaffian at high symmetry points\cite{Kitaev2001,Tewari2012}. As we have only nearest neighbor terms, the invariant in 1D BDI, i.e.~the winding number, only takes the values $\nu=0,\pm 1$. One can also define the parity of a $\mathbb{Z}$ invariant, $\delta=\e^{\im \pi \nu}=\pm 1$. This is generally sufficient for our purposes as the quasi 1D systems in the D class have a $\mathbb{Z}_2$ invariant and in this case $\delta$ and $\nu$ have the same information. For the $\mathbb{Z}$ invariant limited to $\nu=0,\pm 1$, the lost information is small and not crucial to our arguments.

In 2D the Chern number can take several values, though we limit ourselves to plotting only the parity of the invariant in Fig.~\ref{figure1}. For both the 1D and 2D systems the topological phase diagram is independent of the strength of the Rashba coupling, except for the condition that it must be non-zero. In practice, very small Rashba couplings would require very large systems to see clearly defined MBS. As we shall see in the following section, the quasi-1D systems have a much richer phase diagram, which also depends explicitly on the strength of $\alpha_y$. Another example where such dependence on the spin-orbit strength is a hexagonal lattice with a similar Hamiltonian to the one described in Eq.~\eqref{hamiltonian} \cite{Dutreix2014,Dutreix2014b,Sedlmayr2015}. 

Previous models were limited to either an approximate low band theory\cite{Potter2011,Law2011,Stanescu2011,Lim2012,Stanescu2013,Wang2015a} or neglecting the role of spin-orbit coupling all together.\cite{Tewari2012} We go beyond these approximations here by calculating the topological invariant of the Hamiltonian given in Eq.~\eqref{hamiltonian} exactly. This allows a much more accurate depiction of the phase diagram, in particular for large magnetic fields and large spin-orbit coupling where the approximate models do not apply. The experiments on nanowires deposited on the top of superconducting substrates\cite{Mourik2012} are more accurately captured by this model, as it is not in general possible to neglect the role of transverse spin-orbit coupling. 

\section{Topological phase diagrams for quasi-1D wires}\label{sec_topinv}

We will start by computing analytically the exact phase diagram for quasi 1D systems (finite width and infinite length, with open boundary conditions in the transverse direction). For a single wire $N_y=1$, the topological phase diagram is presented in Fig.~\ref{figure1}.  When $N_y>1$ the topological phase diagram can be obtained from a calculation of the topological invariant of the system, as described in what follows.

It is often possible to find a $\mathbb{Z}_2$ invariant for a topological superconductor with particle-hole symmetry by mapping the problem to the parity at the time reversal invariant momenta $(\hat\Gamma_i)$\cite{Sato2009a}.  This is equivalent to calculating the Pfaffian of the Hamiltonian\cite{Kitaev2001,Tewari2012} but proves more convenient for the purposes here. Parity in this case refers to an operation which commutes with the Hamiltonian at these points, $[P,H(\hat\Gamma_i)]=0$, and anti-commutes with the particle-hole operator, $\{P,\C\}=0$. The eigenstates of the Hamiltonian have parity eigenvalues $\pm1$ at the TRI momenta. It follows that there is a representation in which the Hamiltonian  is block diagonal at $\hat\Gamma_i$ with two blocks, one with parity $1$ and the other with parity $-1$. This is the most convenient basis for calculating the invariant, and for a 1D wire or 2D square lattice a trivial re-ordering of the Nambu basis suffices. In the system that we will consider here the problem can be reduced to finding this appropriate basis so that the Hamiltonian is block diagonal at $\hat\Gamma_i$ .

Firstly we note that if $\alpha_y=0$ then the calculation becomes straightforward and standard, see App.~\ref{app_array} or Refs.~\onlinecite{Wakatsuki2014,Poyhonen2014,Seroussi2014,Wakatsuki2014,Deng2014,Sedlmayr2015}. This is because the problem can be decomposed into the transverse momentum channels labelled by the quantum number $k_y$. For each $k_y$ the Hamiltonian at the time reversal invariant (TRI) momenta, $\Gamma_1=0$ and $\Gamma_2=\pi$, is trivially block diagonalizable, exactly as for a 1D wire, in such a way as to define a parity operator which both commutes with the Hamiltonian at the TRI momenta and anti-commutes with the particle-hole operator $C$. All that is required is a re-ordering of the Nambu basis to $\{c^\dagger_{{\vec r}\uparrow},c_{{\vec r}\downarrow},c^\dagger_{{\vec r}\downarrow},-c_{{\vec r}\uparrow}\}$. In this basis the parity operator is simply
\begin{equation}
P_1=\begin{pmatrix}
\mathbb{I}_2&0\\
0&-\mathbb{I}_2
\end{pmatrix}\,.
\end{equation}
Here $\mathbb{I}_n$ is the $n\times n$ identity matrix. This is sufficient for calculating the topological invariant\cite{Sato2009a}.

For the quasi 1D system with transversal Rashba considered here the Hamiltonian is no longer block diagonalizable in any trivial way at the TRI momenta. Nonetheless it is possible to find a suitable parity operator, or equivalently a suitable transformation on the Hamiltonian, by using the transverse spatial, spin and particle-hole subspaces.

After a Fourier transform along the wire the Hamiltonian can be written as $H=\sum_{ k}\Psi^\dagger_{ k}\mathcal{H}( k)\Psi_{ k}$ with
\begin{equation}\label{kham}
\mathcal{H}( k)=\begin{pmatrix}
{\bm f}_{ k}+B & \CuL_{ k} & -\Delta & 0\\
\CuL^\dagger_{ k}& {\bm f}_{ k}-B & 0 & -\Delta\\
 -\Delta  &0 & B-{\bm f}^\dagger_{- k} & \CuL^T_{- k} \\
0&  -\Delta &  \CuL^*_{- k}& -{\bm f}^\dagger_{- k}-B
\end{pmatrix}\,,
\end{equation}
where the entries are themselves matrices for the transverse spatial direction $y$. The terms $B$ and $\Delta$ are diagonal in this space and
\begin{equation}
{\bm f}( k)=\begin{pmatrix}
f( k) & -t & 0 & 0& \ldots\\
 -t& f( k)  &  -t & 0& \ldots\\
 0 &  -t & f( k)  & -t& \ldots\\
  0 & 0 & -t& f( k) & \ldots\\
  \vdots&  \vdots& \vdots&  \vdots&\ddots
\end{pmatrix}\,,
\end{equation}
with
\begin{equation}
\CuL_{ k}=\begin{pmatrix}
\CL_{ k} & -\im\alpha & 0 & 0& \ldots\\
\im\alpha& \CL_{ k} &  -\im\alpha & 0& \ldots\\
 0 &  \im\alpha& \CL_{ k}  & -\im\alpha& \ldots\\
  0 & 0 & \im\alpha& \CL_{ k} & \ldots\\
  \vdots&  \vdots& \vdots&  \vdots&\ddots
\end{pmatrix}\,.
\end{equation}
Finally $f( k) =-2t_x\cos[k]-\mu$ and $\CL_{ k} =-2\im\alpha_x\sin[k]$.

The transformation that allows for the Hamiltonian to be written in block diagonal form at the TRI momenta is $\tilde{\mathcal{H}}( k)=\mathcal{U}^\dagger\mathcal{H}( k)\mathcal{U}$ where
\begin{equation}
\mathcal{U}=\frac{1-{\bm\sigma}^y{\bm\tau}^y}{2}\mathbb{I}_{N_y}+\frac{{\bm\sigma}^z{\bm\tau}^z-{\bm\sigma}^x{\bm\tau}^x}{2}\bar{\mathbb{I}}_{N_y}\,,
\end{equation}
where we introduced an $N_y\times N_y$ matrix whose elements are given by $\left[\bar{\mathbb{I}}_{N_y}\right]_{nn'}=\delta_{n,N_y+1-n'}$. Then
\begin{equation}
\tilde{\mathcal{H}}(\hat\Gamma_i)=\begin{pmatrix}
\bar{\mathcal{H}}(\hat\Gamma_i)&0\\
0&-\bar{\mathcal{H}}(\hat\Gamma_i)
\end{pmatrix}\,.
\end{equation}
In this basis the parity is
\begin{equation}
P_{N_y}=\begin{pmatrix}
\mathbb{I}_{2N_y}&0\\
0&-\mathbb{I}_{2N_y}
\end{pmatrix}\,,
\end{equation}
and we have both $[P_{N_y},\tilde{\mathcal{H}}(\hat\Gamma_i)]=0$ and $\{P_{N_y},\C\}=0$. We note that also $[\C,\mathcal{U}]=0$.

\begin{figure}
\includegraphics*[width=0.95\columnwidth]{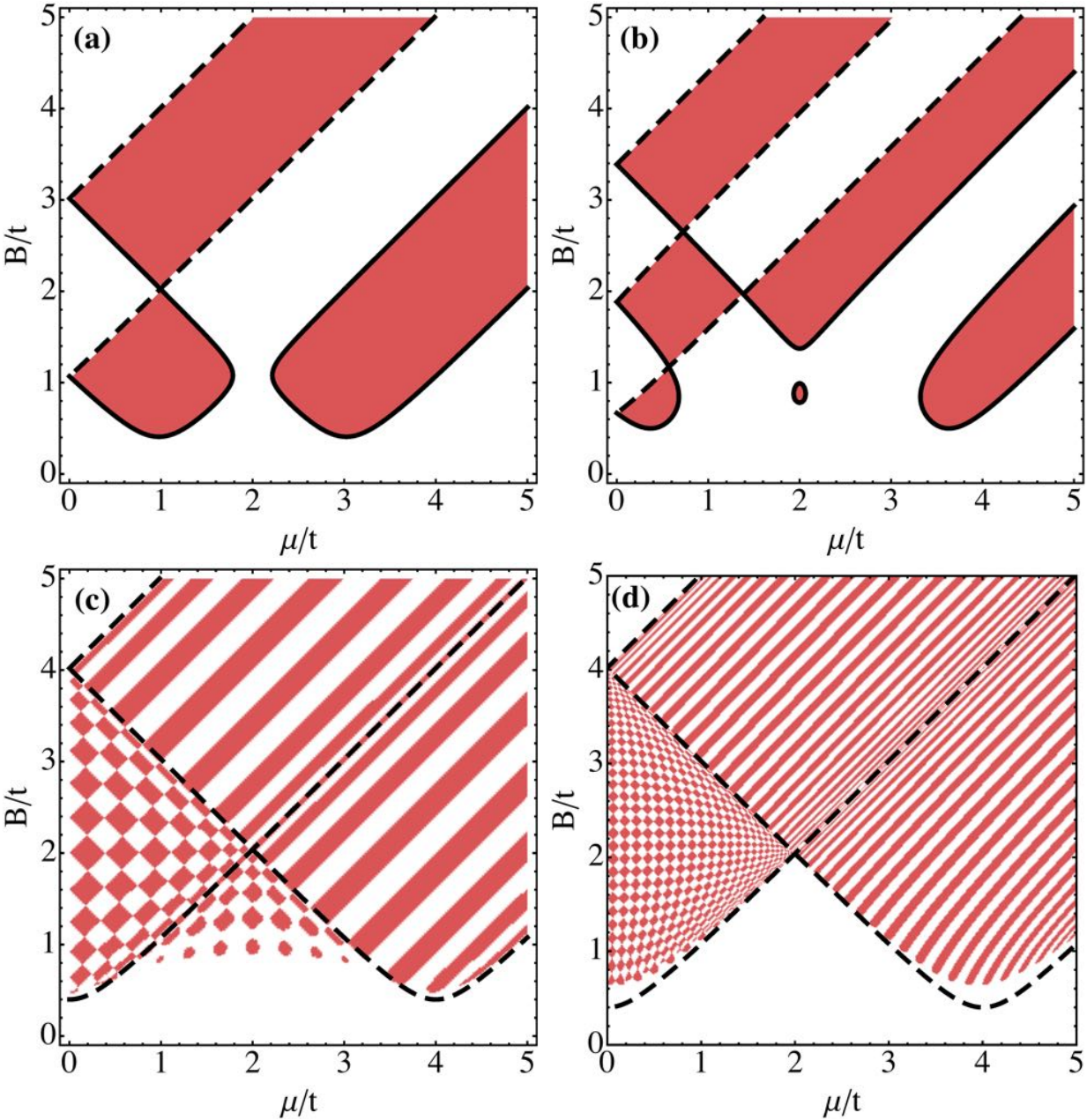}
\caption{(Color online) Topological phase diagram as a function of $B$ and $\mu$ for a ladders with (a) $N_y=2$, (b) $N_y=3$, (c) $N_y=15$, and (d) $N_y=50$ with $\Delta=0.4$. Light red is the topologically non-trivial phase and white is topologically trivial. For (a,c) $\alpha_x=\alpha_{y}=0.2$ and for (b,d) $\alpha_x=\alpha_y=0.5$. For (a,b) gap closing lines belong to the two different TRI momenta are shown as dashed and full lines . In (c) and (d) the gap closing lines for  the 2D bulk system are shown for comparison.}
\label{figure2}
\end{figure}

The topological invariant is then calculated in the usual way\cite{Sato2009}
\begin{equation}\label{invariant}
\delta=\sgn\det\bar{\mathcal{H}}(\hat\Gamma_1)\det\bar{\mathcal{H}}(\hat\Gamma_2)\,.
\end{equation}
When $\delta=-1$ there is band inversion, i.e.~the parity switches between the TRI momenta an odd number of times and the system is topologically non-trivial. For $\delta=1$ the system is topologically trivial, see Fig.~\ref{figure2} for some examples.

The full analytical expressions become quickly cumbersome as $N_y$ is increased. Full expressions for $N_y=2,3$ are give in App.~\ref{app_inv}.
One key point to note is that because of the transverse spin orbit coupling the expressions $\det\bar{\mathcal{H}}(\hat\Gamma_{1,2})$ do not factorize further into simple expressions for each wire.

The two wire system (two-leg ladder) presented in Fig.~\ref{figure2}(a) has a peculiar property: close to the point $B=2t$ and $\mu=0$ inside the topologically trivial phase, two pairs of MBS can form. These states are not however topologically protected and can be destroyed by introducing disorder\cite{Sedlmayr2015a}. The zero overlap of these MBS is sometimes referred to as a ``hidden symmetry''\cite{Dumitrescu2015b}. 

In contrast to a wire or square lattice, but similar to a hexagonal lattice\cite{Dutreix2014b,Sedlmayr2015}, the topological phase diagram depends explicitly on all of the parameters including the Rashba spin-orbit coupling.
Thus
we note that the number of the topological regions in the area close to $\mu=2$, and $B<2$ (between the 1D and 2D transition lines) depends on both the width of the system, as well as on the value of the Rashba coupling, and it strongly reduced for large Rashba values as well as for a large $N_y$. One must thus note that switching on the spin-orbit coupling while keeping the other parameters constant, can also drive the system through a topological phase transition.  Also, while in this section we focus mostly on the case of isotropic coupling, anisotropic systems in which the spin-orbit interaction and hopping strengths may not be symmetric between the $x$ and $y$ directions are explored in App.~\ref{app_asymm}. We note that for finite $\alpha_x$, when $\alpha_y\to0$ we recover the phase diagram for an array of wires, see Fig.~\ref{figure8}. In this limit the nanowire can host many topologically protected Majorana bound states. Switching on $\alpha_y$, which breaks a time reversal symmetry of the problem, gaps out all but at most two of the MBS. Naturally in the weak coupling limit it is possible to infer the topological phase diagram from that for $\alpha_y=0$\cite{Tewari2012}.

Note the evolution from the phase diagram for a single wire shown in Fig.~\ref{figure1} to the complex pattern that we obtain for a large $N_y$. It is important to stress here the difference between the phase diagram for a system with open boundary condition in which $N_y$ is very large (such as Fig.~\ref{figure2}(d) ), and the bulk phase diagram for the 2D system depicted in Fig.~\ref{figure1}(b). The latter corresponds to the topological phase diagram describing the formation of Majorana edge state for a system with infinite $N_y$ and in which periodic boundary conditions are imposed in the $y$ direction. The main difference consists in the fact that in the system with open boundary conditions the phase diagram remains discreet, and the trivial and topological phases alternate with $\mu$ and $B$ with periodicities which become smaller and smaller with increasing the width of the system. The source of this discreteness is the fact that, no matter the size of the system, it is still an open system with a single boundary. The open system, invariant of dimension, is fundamentally different from the infinite nanoribbon (periodic system), and we believe that this difference has not been thoroughly explored, especially in connection with experimental systems such as quasi 1D wires, which should rather be modeled as finite-width open systems. A more detailed discussion of the difference between fully open systems (flakes) and systems with periodic boundary conditions (tubes) as well as the transition between the two cases, is presented in Appendix C.

Note also that the existence of the discrete steps in the phase diagram (known also as the `even-odd' effect) has been also predicted for the multi-band wires\cite{San-Jose2014}: the relation between our exact calculation valid for wires of arbitrary widths and in all the parameter space and the multi-band approximation is explored in detail in Sec.~\ref{sec_multiband}.

\section{2D open system}\label{sec_2D}

In what follows we consider fully open systems of finite dimensions in both directions, and we analyze their properties numerically by looking at the lowest energies eigenstates obtained using the MathQ code \cite{Note1}. If $N_y \gg N_x$ one expects to recover the phase diagrams obtained analytically in the previous section. The interesting question is what happens when the transversal and longitudinal directions become comparable, in which situation we should expect that the system has effectively a single boundary and the Majorana states on the edges would hybridize and destroy each other. The manner in which this happens, as well as the persistence of some topological properties in the low-energy edge states of a finite-size open system are investigated using the generalized Majorana polarization introduced in Ref.~\onlinecite{Sedlmayr2015b}.

\subsection{Phase diagram deduced from the Majorana polarization}

While for an infinite system the condition to be in a topological state can be determined using the topological invariant, such criterion cannot be applied to finite-size systems. This can happen for example because finite-size effects can destroy the Majorana states even in regions in the phase space that are characterized as topological from a bulk perspective.  Thus one needs to apply a different criterion for such systems to test the formation of Majoranas. 
Moreover, solely the existence of zero-energy states is not enough since many non-Majorana states, such as impurity states, can form at zero energy; also many infinitesimally-small-energy states can arise which are hard to distinguish from purely zero-energy states. To test if such states are Majorana or not we are applying here the criterion of the generalized Majorana polarization (MP) \cite{Sedlmayr2015b}. 

The MP is a local measure characterizing a specific wavefunction which can tell if such state is a Majorana or not. While the topological invariant is a characteristic of the `Hamiltonian' of a given system, the MP is a characteristic of the `eigenstates' of the Hamiltonian, as calculated for a specific configuration, for a system of a given size and with particular boundary conditions. Thus, in the same way that the local density of states (LDOS) can provide space-dependent information about the formation of edge states for a given configuration of the system, information which is not solely dependent on the Hamiltonian, and cannot be extracted directly from it, the MP can tell about the formation of Majoranas in a given system by analyzing its eigenvalues and eigenfunctions, and their dependence on position. The MP is thus an equivalent of the LDOS measuring the Majorana density at a given position in space. However, while the LDOS is a scalar quantity, the MP is a vector in the complex plane (we can see it as a Majorana pseudo-spin), whose magnitude describes the Majorana density, i.e. the electron-hole overlap, and its phase the relative phase between the electron and hole components. 

\begin{figure}
\includegraphics*[width=0.95\columnwidth]{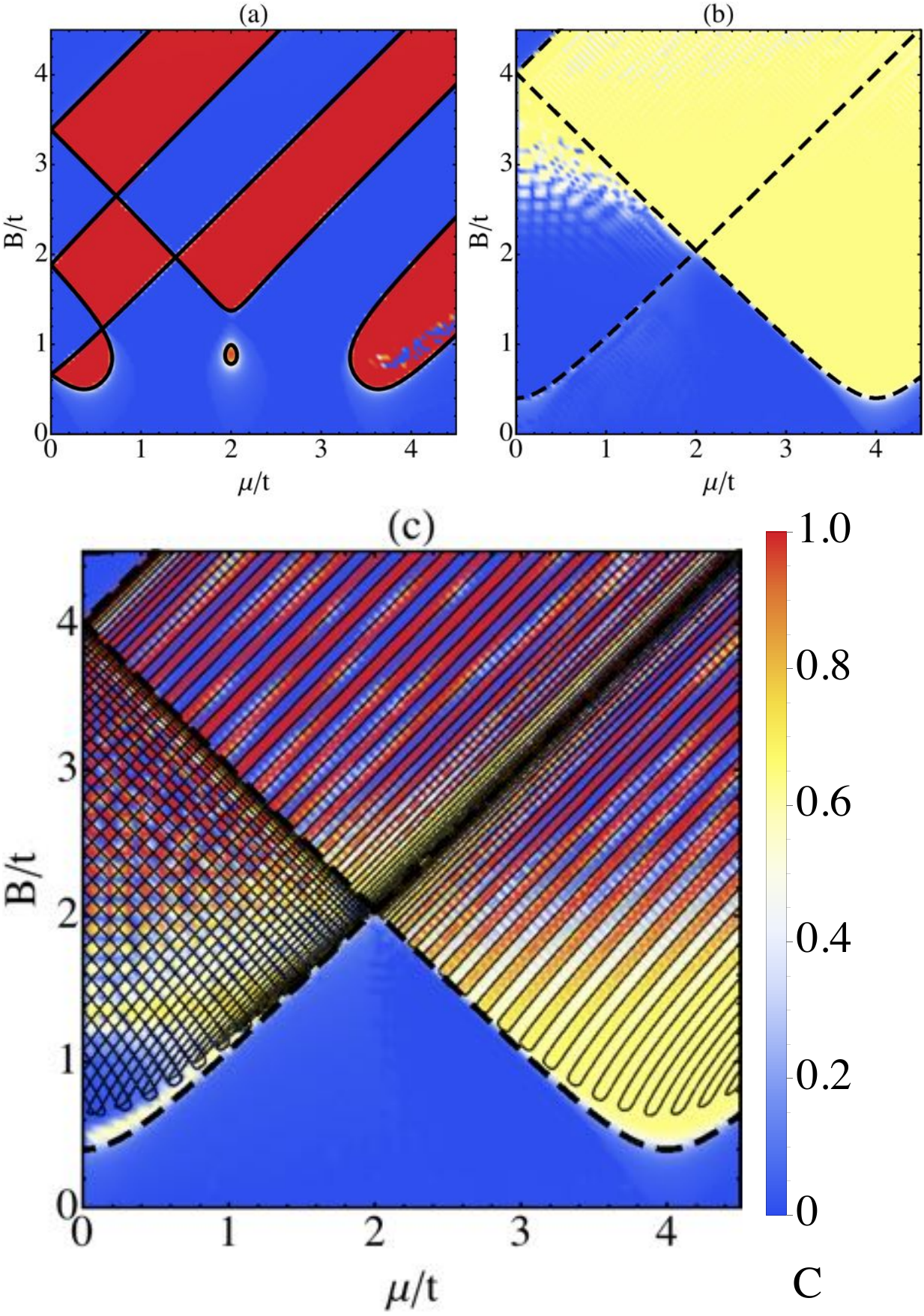}
\caption{(Color online) The phase diagram as a function of $B$ and $\mu$ for open systems of (a) $3\times201$, (b) $101\times101$, and (c) $51\times201$. This has been calculated using the PH expectation value of the lowest energy state summed over half of the system $C$ (see main text). In all the examples $\Delta=0.4t$ and $\alpha=0.5t$. The analytical topological phase boundaries for quasi 1D (solid lines) and 2D (dashed lines) are also plotted.}
\label{figure3}
\end{figure}

The way in which the MP criterion works is as follows\cite{Sedlmayr2015b}: a Majorana state is an eigenstate of the particle hole operator. Therefore a Majorana-like state localized inside a spatial region $\mathcal{R}$ must satisfy $C=1$ where $C$ is the magnitude of the integral of the Majorana polarization vector over the spatial region $R$.
\begin{equation}
C=\frac{\left|\sum_{{\vec r}\in \mathcal{R}}\langle\Psi|\C{\hat r}|\Psi\rangle\right|}{\sum_{{\vec r}\in \mathcal{R}}\langle\Psi|{\hat r}|\Psi\rangle}\,.
\end{equation}
${\hat r}$ is the projection onto site ${\vec r}$, and the local MP is simply the expectation value of the local particle-hole transformation:
\begin{equation}
\langle\Psi|\C{\hat r}|\Psi\rangle=-2\sum_\sigma\sigma u_{{\vec r}\sigma}v_{{\vec r}\sigma}\,.
\label{mp1}
\end{equation}
We split our system in half such that  $\mathcal{R}$ is half of the system (divided usually along the longer length).  The Majorana polarization also allows us to distinguish between zero energy states which are Majorana bound states and those which are not as $C=1$ is both a necessary and sufficient condition for a state being a Majorana state.
By numerically solving the tight-binding Hamiltonian, Eq.~\eqref{hamiltonian}, and plotting $C$ as a function of the parameters we can accurately recover the appropriate topological phase diagram. The topological phases are characterized by $C=1$, while the non-topological ones by $C=0$. This is demonstrated in Fig.~\ref{figure3} for a few systems of different sizes.

The MP thus allows an accurate characterization of the topological phase even in cases where a direct calculation of the topological index is not feasible. Moreover it does not require very large system sizes to be highly accurate, as for example a map of the lowest energy state would. In addition, as we will shortly demonstrate, the MP gives local information about the formation and behavior of the Majorana bound states which can not be gained form a consideration of the system's bulk properties. Thus we should note that, while if a Majorana state exists the system is surely topological, the reverse is not necessary true, as a bulk topological system may have no Majorana solutions in certain finite-size configurations.

For a system in which the length is much larger than the width $N_y\ll N_x$ such as the one depicted in Fig.~\ref{figure3}(a) we would expect to recover the phase diagram calculated analytically in the previous section, and this is indeed the case, the topological regions predicted by the MP criterion (red in Fig.~\ref{figure3}(a)) correspond exactly to the topological regions (light red) in Fig.~\ref{figure2}(b). For a wider system such as the one depicted in Fig.~\ref{figure3}(c) we can see that the phase diagram calculated analytically (Fig.~\ref{figure2}(d)) is recovered very well at large values of B, however, at small values of B, the value of $C$ is neither $0$, nor $1$, but it shows an intermediate value (depicted in yellow). For fully square systems (Fig.~\ref{figure3}(b)) the entire topological phase is characterized now by $C\approx 0.7$. We should note that all the states corresponding to the regions with $C<1$ have non-zero but very small energies, thus making it harder to interpret them as Majoranas or not based solely on the energy criterion. 

\begin{figure}
\includegraphics*[width=0.95\columnwidth]{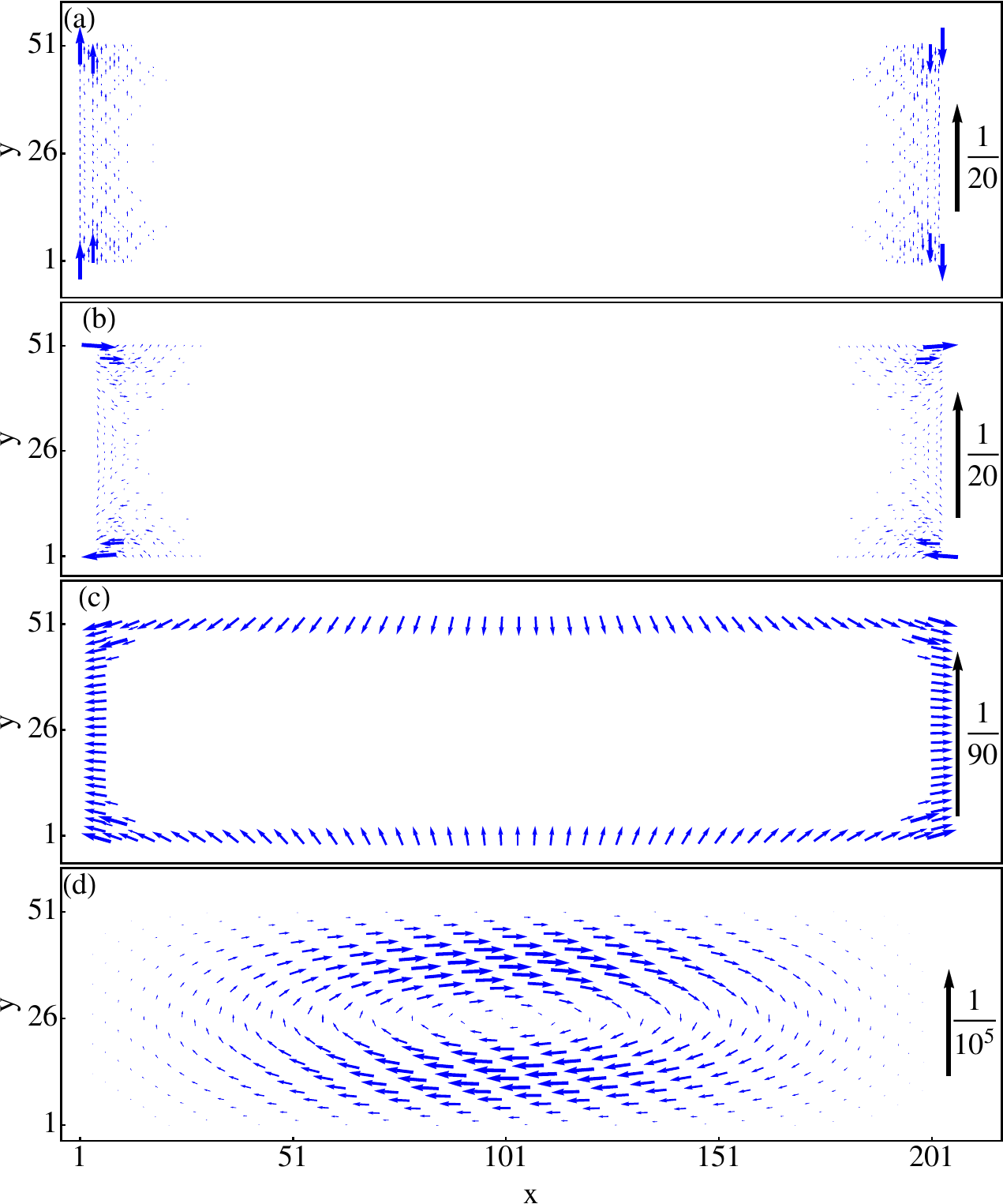}
\caption{(Color online) MP for the lowest energy states for open system of $51\times201$ with $\Delta=0.3t$ and $\alpha=0.5t$.  We plot (a) a MBS $\mu=3.5t$ and $B=2.2t$ (corresponding to a red phase in Fig.~\ref{figure3}(c)), (b) a 1D edge state $\mu=3.5t$ and $B=2.3t$ (blue phase in Fig.~\ref{figure3}(c)), (c) a 2D edge state $\mu=3.5t$ and $B=t$ (yellow phase in Fig.~\ref{figure3}(c)), and (d) a bulk state $\mu=0.5t$ and $B=5t$ (blue phase in Fig.~\ref{figure3}(c)).}
\label{figure4}
\end{figure}

\begin{figure}
\includegraphics*[width=0.95\columnwidth]{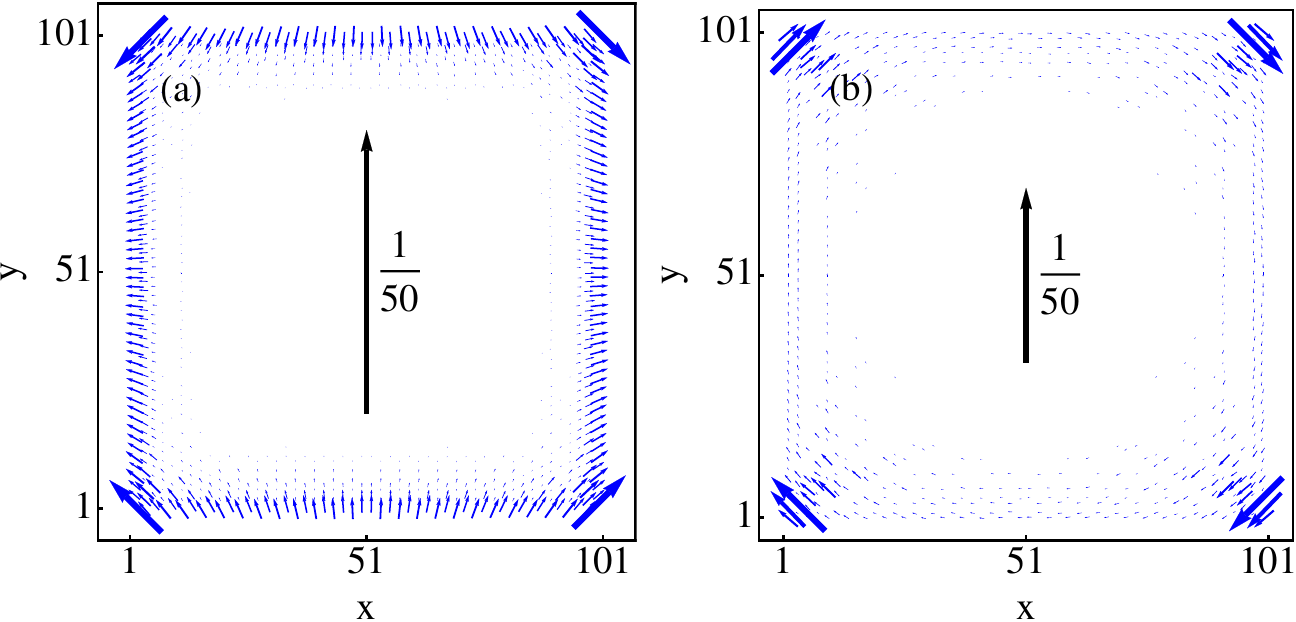}
\caption{(Color online)  MP for lowest energy state for open system of $101\times101$ with $\Delta=0.3t$ and $\alpha=0.5t$.  (a) $\mu=4t$ and $B=t$, and (b) $\mu=2t$ and $B=3t$, both in the 2D topologically non-trivial phase.}
\label{figure5}
\end{figure}

To understand these states we focus on the local structure of the MP defined in Eq.~(\ref{mp1}) as a vector in the complex plane. A Majorana state must have an integral of the MP of $1$ over a spatial region $\mathcal{R}$, thus it must exhibit an uniform phase inside this region, equivalently the MP local vectors need to be aligned inside $\mathcal{R}$ (`ferromagnetic' MP structure).
In Figs.~\ref{figure4}-\ref{figure5} we plot the MP for a variety of different low energy states. In Fig.~\ref{figure4} we plot the MP vector for a  $51\times201$ system with $\Delta=0.3t$ and $\alpha=0.5t$. Fig.~\ref{figure4}(a) corresponds to the red (topological) phase in Fig.~\ref{figure3}(c) ($C=1$), with $\mu=3.5t$ and $B=2.2t$. We should expect to have two MBS confined at the two narrow ends of the ribbon, and we note that this is indeed the case,  we can see the formation of two `ferromagnetic' states localized at the two ends of the wire, with opposite MP. Fig.~\ref{figure4}(b) corresponds to a non-topological (blue phase in Fig.~\ref{figure3}(c)) phase, with $C=0$. We take $\mu=3.5t$ and $B=2.3t$, and the corresponding state, while being localized on the edges, is indeed non-Majorana, which can be seen from the local fluctuations of the direction of the MP vector from site to site, which makes its integral over half of the wire zero.
In Fig.~\ref{figure4}(c) we focus on the more puzzling $C<1$ phase, denoted in yellow in Fig.~\ref{figure3}(c). We take $\mu=3.5t$ and $B=t$. We note that the corresponding lowest energy state is also an edge state, extending over the entire contour of the system. Moreover, while the state is not `ferromagnetic', a small uniform variation is observed from site to site, making the total integral over $R$ finite, but not equal to $1$. 
 Finally in Fig.~\ref{figure4}(d) we show a state with $C=0$ for $\mu=0.5t$ and $B=5t$, which corresponds to the blue region outside the region delimited by the bulk 2D topological lines, this actually correspond to a bulk state for which the value of the MP is small even locally. 

It appears that the existence of the `yellow' phase is due to the shape of the system, i.e to the fact that the length and the width become comparable. If the length is increased the yellow phase is diminishing and the phase diagram converges towards the phase diagram calculated analytically in the previous section. This observation has been made also for the Kitaev model \cite{Sedlmayr2015b}. However, we want to stress that in real experimental systems the existence of the intermediate phase with $C<1$ is possible, and it would correspond to the formation of low-energy subgap states which have locally a full Majorana character, but that are non-Majoranas, due to their global lack of symmetry. It would be interesting to explore the connection between these states and the 'chiral' Majorana states described in Refs.~\onlinecite{Potter2010,Alicea2012}. Another interesting question would be if such quasi-Majorana states have any topological characteristics, such as non-Abelian statistics, or atypical braiding properties, and if such states can be useful for example for quantum computation in a similar manner as the Majorana states. 
Note that a state with a $C=1$ is a Majorana, and certainly has non-Abelian statistics. It is not at all clear that the statistics of the quasi-Majorana states would be Abelian. These are peculiar states in-between fermionic and Majorana-like, and there exist to this point no work to investigate their statistics. Moreover there seems to be a fundamental difference between the properly fermionic states which have a uniform local zero MP. The quasi-Majoranas have their weight divided into a few different regions, roughly disconnected from each other, each with $C$ different from zero. Manipulating such a state in one part of the system is not the same as manipulating the fermionic state, and the effect of such manipulation is also a very interesting question. Such questions are however well beyond the scope of the present work.

An extreme situation is depicted in Fig.~\ref{figure5}(a,b), in which we plot the MP for two quasi-Majorana states for a square system. Fig.~\ref{figure5}(a) depicts the MP for a set of parameters inside the 2D bulk topological phase (as described in Fig.~\ref{figure1}(b)), while Fig.~\ref{figure5}(b) describes a set of parameters inside the 1D topological phase (as described in Fig.~\ref{figure1}(a)), that would become a Majorana end state if one direction of the system was sufficiently increased. Interestingly enough these two states have different MP characteristics, the `1D' one being localized roughly at the corners, while the `2D' one along the contour of the system, allowing one to distinguish between the two. Note that for a fully square system the corner MP vectors are oriented at roughly $90$ degrees angle with respect with each other, giving rise to an integral of $C=\sqrt{2}/2\approx 0.7$, which corresponds well to the `yellow' in Fig.~\ref{figure3}(b).

\section{Local density of states}\label{sec_dos}

It would be interesting to see the effects of the discrete topological transitions in the phase diagram of the open systems on directly observable quantities such as the local density of states (LDOS) measurable in STM or tunneling experiments. This has been explored for systems with a small number of bands in certain regions of the parameter space\cite{Stanescu2011,Stanescu2013,Wakatsuki2014,San-Jose2014,Dumitrescu2015a}, but in order to make contact with the experiments we need to explore systems of different sizes and geometries. In what follows we focus on both quasi-1D systems, as well as on systems in which the width becomes comparable to the length of the wire.
The LDOS at one end of the wire is defined as
\begin{equation}
\rho(\varepsilon)=\sum_n\left\langle\psi_{n}\left|({\bm\tau}^0+{\bm\tau}^z){\hat r}_{\rm edge}\right|\psi_{n}\right\rangle\frac{\e^{-\frac{(\varepsilon-\epsilon_n)^2}{\Gamma^2}}}{2\sqrt{\pi}\Gamma}\,.
\end{equation}
$\Gamma$ is a broadening energy scale, typically taken to be slightly larger than the mean level spacing. To avoid effects due to local fluctuations we define the edge LDOS as an average over all sites on the first $5$ atomic rows away from the edge.

\begin{figure}
\includegraphics*[width=0.98\columnwidth]{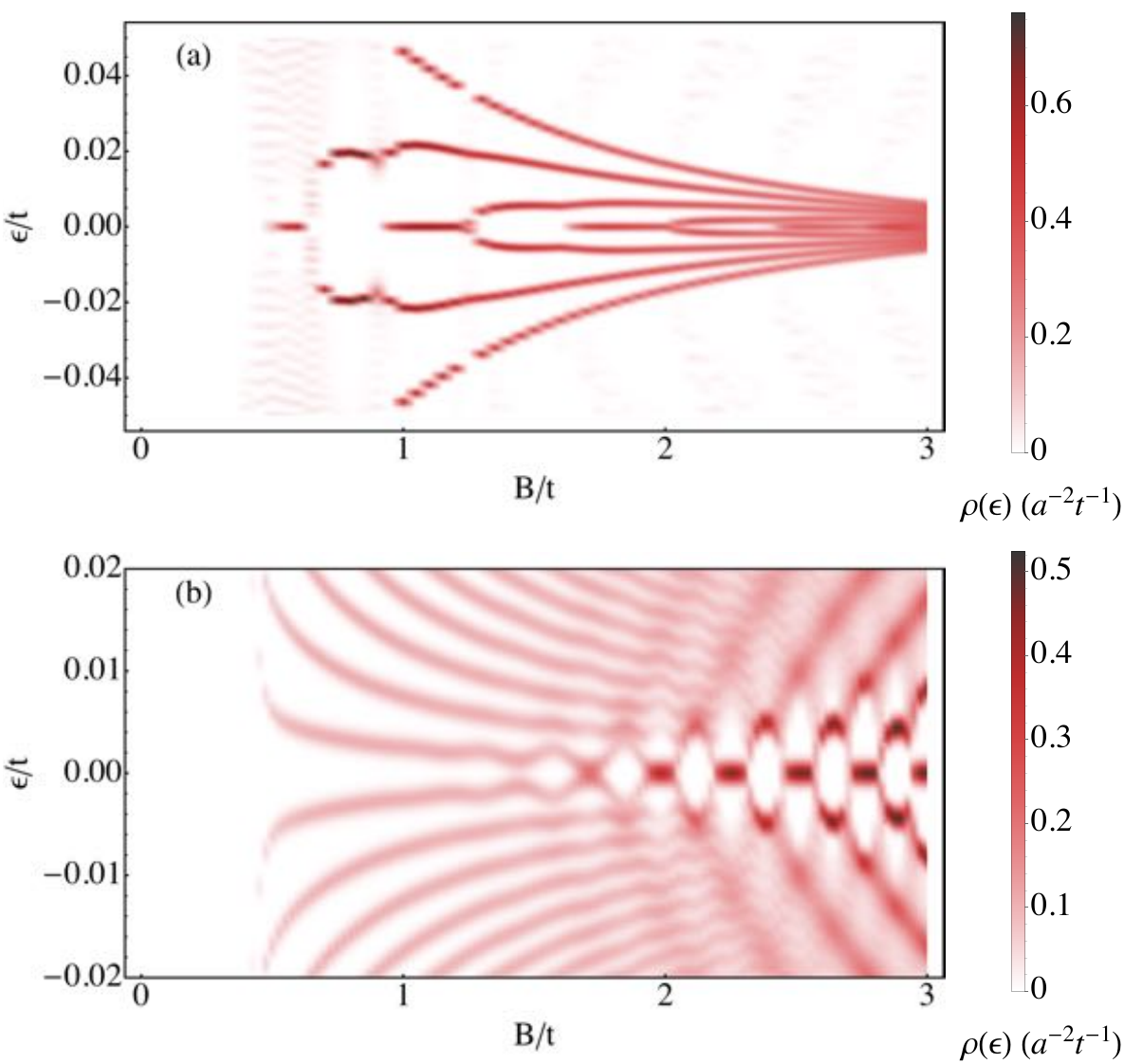}
\caption{(Color online) The local density of states at the edge of quasi-1D wires as a function of energy and $B$ with $\Delta=0.4t$, $\mu=4t$, and $\alpha=0.5t$. Panel (a) is for $N_y=15$ and (b) is for $N_y=51$ , in all cases $N_x=201$.}
\label{figure6}
\end{figure}

In Fig.~\ref{figure6} the LDOS at one end of the quasi-1D wire is shown for two examples, (a) $15\times 201$ sites and (b) $51\times 201$ sites. In both examples $\Delta=0.4t$, $\mu=4t$, and $\alpha=0.5t$. The bulk gap closing is only faintly visible visible in Fig.~\ref{figure6}(a) due to the small weight of the bulk states at the edges of the system. 
We note the formation of MBS edge states inside the parameter regions predicted by the topological phase diagram. In addition, we note that even  in the topologically trivial phases, the lowest energy states are also edge states, even if split in energy and non-Majoranas (see also the characterization of such a `blue' state in the previous section). 
Note that these edge states can be understood as MBS of a BDI model split by the transverse Rashba, see App.~\ref{app_array}.  As the strength of the Rashba coupling is relatively small in the nanowires currently used\cite{Mourik2012} this splitting could be experimentally very small.

\section{Effective multi-band theory}\label{sec_multiband}

For certain ranges of parameters the phase diagram for a fully open system can be calculated using a multi-band approximation. Such an approximation is valid if the width $N_y$ is large, respecting however the quasi 1D condition, $N_x\gg N_y$\cite{Potter2010,Potter2011,Stanescu2011,Stanescu2013}. However it is clear that much of the rich physics that we can find exactly for system of arbitrary width and length cannot be recovered using this approximation. In what follows we show how one can recover the appropriate multi-band limit for our model, based on a truncation of the active bands in the model, and we show how it compares to the exact result in terms of the regions of validity.

Let us start by relating our tight-binding model to a continuum model with a quadratic kinetic energy: $-(\partial^2_{xx}+\partial^2_{yy})/2m$, with $m$ being the effective mass. Discretizing on a lattice scale $a$ we find the lattice Hamiltonian Eq.~\eqref{hamiltonian} with a corresponding hopping $t=1/(2ma^2)$ and a scaled chemical potential of $\delta\mu=\mu-\mu_0$, and $\mu_0=4t$. Thus $\delta\mu=0$ corresponds to the point where we can define our effective model which is the bottom of the band.

To begin we substitute
\begin{equation}
\psi(x,y)=\sum_{n=1}^{N_y}\tilde\psi_{nx}\sqrt{\frac{2}{N_y+1}}\sin\frac{\pi n y}{N_y+1}
\end{equation}
into Eq.~\eqref{hamiltonian}. Then we have
\begin{eqnarray}\label{hamq1d}
 H_{\rm Q1D}&=&\sum_{n,x}\bigg[ \tilde\psi^\dagger_{nx}\left(\epsilon_{n}{\bm\tau}^z-\delta\mu{\bm\tau}^z-\Delta{\bm\tau}^x+B{\bm\sigma}^z\right)\tilde\psi_{nx}\nonumber
\\&&+\tilde\psi^\dagger_{nx}\left(-t-\im\alpha_x{\bm\sigma}^y\right){\bm \tau}^z\tilde\psi_{n,x+1}+\textrm{H.c.}\bigg]
\\&&-\sum_{\substack{x\\nn'}}\tilde\psi^\dagger_{nx}\frac{2\im\alpha_y}{N_y+1}{\bm\sigma}^x{\bm\tau}^zM_{nn'}\tilde\psi_{n'x}\,,\nonumber
\end{eqnarray}
where $\epsilon_n=-2t\cos[k_n]$, $k_n=\pi n/(N_y+1)$ and
\begin{equation}
M_{nn'}=\begin{cases}0\hspace{3.4cm}\textrm{if $n+n'$ is even,}\\
\frac{\sin[(k_n+k_{n'})/2]}{\sin[(k_n-k_{n'})/2]}-\frac{\sin[(k_n-k_{n'})/2]}{\sin[(k_n+k_{n'})/2]}\textrm{ otherwise.}\end{cases}
\end{equation}
By truncating the transverse modes at some particular $n^*$ we obtain an approximate theory consisting of just $4n^*$ bands rather than $4 N_y$. We will refer to $n^*$ as the number of channels, and each channel carries both spin and particle-hole degrees of freedom. The energy difference $\ell_{n^*}\equiv\epsilon_{n^*}-\epsilon_1$ gives an energy scale for the multi-band theory. We need to impose $B,\Delta,\alpha\ll\ell_{n^*}$, where for large wire widths $(N_y\gg 1)$ we can write
\begin{equation}
\ell_{n^*}\approx\frac{\pi^2 t}{N_y^2}\left[\left(n^*\right)^2-1\right]\,.
\end{equation}
Naturally the mixing between the bands with $n\leq n^*$ and $n>n^*$ must be negligible compared to $\epsilon_{n^*}$, i.e.~the Rashba coupling must be small.

In order to calculate the topological invariant we proceed exactly in the same way as for the exact calculations presented in Sec.~\ref{sec_model}, and we start by performing a Fourier transform along the bulk direction of the wire. The Hamiltonian becomes $H=\sum_{ k,nn'}\Psi^\dagger_{kn}\mathcal{H}_{nn'}(k)\Psi_{kn'}$ with $\mathcal{H}_{nn'}(k)=\mathcal{H}^0_n(k)\delta_{nn'}+\mathcal{H}^{y}_{nn'}$ and
\begin{equation}\label{kham0}
\mathcal{H}^0_n( k)=\begin{pmatrix}
f_{kn}+B & \CL_{ k} & -\Delta & 0\\
\CL^\dagger_{ k}& f_{ kn}-B & 0 & -\Delta\\
 -\Delta  &0 & B-f^\dagger_{- k,n} & \CL^T_{- k} \\
0&  -\Delta &  \CL^*_{- kn}& -f^\dagger_{- k,n}-B
\end{pmatrix}\,.
\end{equation}
where $f_{kn} =-2t\cos[k]-2t\cos[k_n]-\delta\mu$ and $\CL_{ k} =-2\im\alpha_x\sin[k]$. The band mixing terms are
\begin{equation}\label{khamy}
\mathcal{H}^y_{nn'}=-\frac{2\im\alpha_y}{N_y+1}\begin{pmatrix}
0 & M_{nn'} & 0 & 0\\
M_{nn'}&0& 0 &0\\
0  &0 &0 & M_{nn'} \\
0&  0 &  M_{nn'}& 0
\end{pmatrix}\,,
\end{equation}

The transformation that allows for the Hamiltonian to be written in block diagonal form at the TRI momenta is  $\tilde{\mathcal{H}}_{nn'}(k)=\mathcal{U}^\dagger_n\mathcal{H}_{nn'}(k)\mathcal{U}_{n'}$ where
\begin{equation}
\mathcal{U}_n=\frac{1-{\bm\sigma}^y{\bm\tau}^y}{2}+(-1)^n\frac{{\bm\sigma}^z{\bm\tau}^z-{\bm\sigma}^x{\bm\tau}^x}{2}\,.
\end{equation}

\begin{figure}
\includegraphics*[width=0.95\columnwidth]{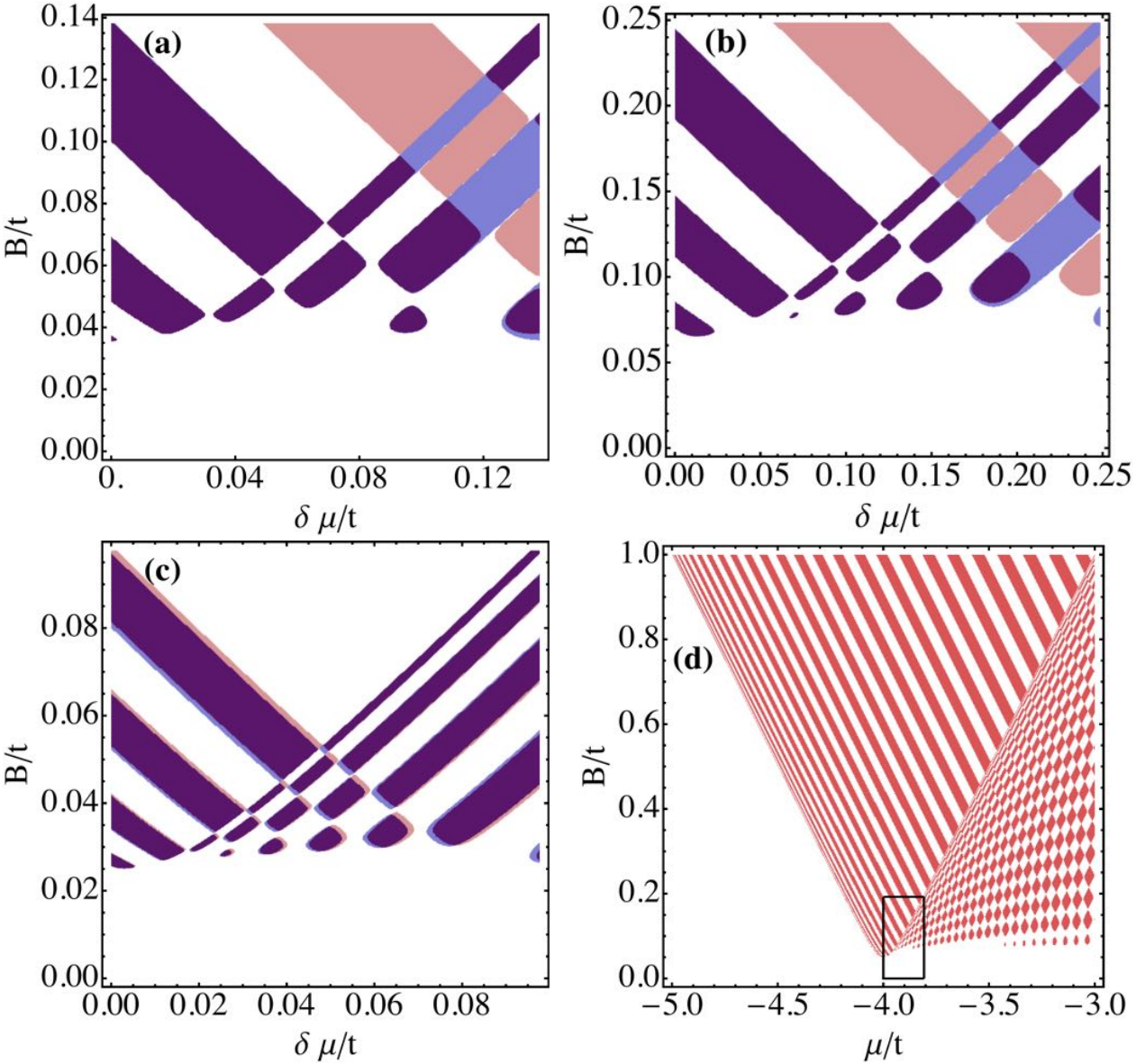}
\caption{(Color online) A comparison of the exact and effective phase diagrams as a function of $B$ and $\mu$. We take $\Delta=0.25\ell_{n^*}$ and $\alpha=0.1\ell_{n^*}$. In panels (a,b) we present a comparison of the full and multi-band effective topological phase diagram for $N_y=50$ and $n^*=6,8$ respectively. The white (topologically trivial) and dark purple (topologically non-trivial) phases denote the regions where both calculations agree. The pink regions are predicted to be topological by the exact calculation and non-topological by the multi-band approximation, while the light purple regions are predicted to be non-topological by the exact calculation and topological by the multi-band approximation. Panel (c) shows a comparison between the phase diagrams obtained by the multi-band approximation with $n^*=10$, for systems with $N_y=100$, and $N_y=101$ respectively. The small pink regions are the extra-topological regions corresponding to $N_y=101$. Panel (d) shows the exact topological phase diagram for $N_y=100$ with the region where the multi-band theory is working (corresponding to the results plotted in (c))  boxed by the small rectangle
.}
\label{figure7}
\end{figure}

Following the same procedure as the one presented in Sec.~\ref{sec_model} recovers the phase diagram more accurately. In Fig.~\ref{figure7}(c) we compare the approximate phase diagrams for systems with different numbers of wires, $N_y=100$ and $N_y=101$, obtained using the same number of channels, $n^*=10$. The sole difference between the two phase diagrams are the small extra topological regions denoted in pink for $N_y=101$ which are barely visible, confirming that the multi-band theory is not very sensitive to $N_y$ in its region of validity. Last, but not least, in Fig.~\ref{figure7}(d) we plot the exact phase diagram for a system with $N_y=100$, with the small boxed region corresponding to the region of validity for the multi-band approximation, indicating that only a very small part of the full phase diagram can be recovered using this approximation.

We want to emphasis that although it is always possible to construct an effective multi-band theory, except for the cases where only 1 or 2 channels are sufficient to describe the systems, which are indeed more tractable analytically, the exact model has in general quite an advantage of use over the multi-band approximation. This is because it not only captures accurately the physics for a wider parameter range, but it also allows in general for a more straightforward calculation of the topological invariant. This is simply because the real space provides a simpler description of the transverse direction in the presence of both the spin-orbit coupling and of the open boundaries.

\section{Conclusions}\label{sec_conclusions}
We have studied the formation of Majorana bound states in square lattices with longitudinal and transversal Rashba, a Zeeman magnetic field, and in the presence of superconducting proximity. We have considered fully open systems, which we believe is the most relevant situation for experiments. We have first calculated exactly analytically the topological phase diagrams for long quasi 1D wires using a topological invariant. We have shown that small regions of this phase diagram can be reproduced by a multi-band approximation under certain conditions and for certain system parameters. We have then explored numerically the phase diagrams for fully open systems with arbitrary dimensions and geometries using the Majorana polarization. We have found that for very elongated systems the MP criterion recovers the phase diagrams obtained analytically. For systems in which the length and the width of the system become comparable we find that more peculiar features arise: thus the low energy states are no longer Majoranas but quasi-Majorana states, i.e. states that have a small but non-zero energy, as well as locally a perfect electron-hole overlap, but no global Majorana symmetry. For square systems no true Majorana states can form but the quasi-Majorana states form in the entire region corresponding  to the bulk 2D topological phase. We have also examined the asymmetry between transversal and longitudinal Rashba, as well the transition between the periodic and open boundary conditions.

It would be interesting to explore these quasi-Majorana states more thoroughly, in particular, given the likelihood to generate such low-energy states in experimental systems, in would be useful to investigate their possible interest for braiding operations and quantum computation as well as their statistics. Also, it would be interesting to understand the relation between these states and the chiral Majorana states \cite{Potter2010,Alicea2012}.

It would be also interesting to generalize our work to quasi 1D wires with a three-dimensional structure and with and more complicated lattice structures and tight-binding models,as well as to explore the modification of the direction of the magnetic field, to see if the discrete phase transitions in the phase diagram can reproduce the experimental data for InAs and InSb wires\cite{Mourik2012}.

\acknowledgements

This work is supported by the ERC Starting Independent Researcher Grant NANOGRAPHENE 256965. We would like to thank Pablo San-Jose for extensive help in implementing the MathQ code, as well as for detailed discussions and comments about our work and the manuscript. We also acknowledge interesting discussions with Pascal Simon, Denis Chevallier, and Mircea Trif.

\appendix

\section{Analytical expressions for the topological invariant for: $N_y=2,3$}\label{app_inv}

We give here the full expression for the topological invariant for $N_y=2,3$. For larger numbers of wires it becomes quickly unwieldy, though it can always be calculated using the method we have shown, even up to very large numbers of wires. From Eq.~\eqref{invariant} in the main text we have the topoloigcal invariant
\begin{equation}\label{invariantapp}
\delta=\sgn\det\bar{\mathcal{H}}(\hat\Gamma_1)\det\bar{\mathcal{H}}(\hat\Gamma_2)\,.
\end{equation}
For $N_y=2$ one finds, reinserting $t_x=t_y=t$ for clarity and defining $f_{k}=-2t\cos k-\mu$,
\begin{widetext}
\begin{equation}\label{ny2h}
\bar{\mathcal{H}}(\hat\Gamma_{1,2})=
\begin{pmatrix}
-t&f_{0,\pi}+B&0&-\im\alpha_y+\Delta\\
f_{0,\pi}+B&-t&\im\alpha_y+\Delta&0\\
0&-\im\alpha_y+\Delta&t&-f_{0,\pi}+B\\
-\im\alpha_y+\Delta&0&-f_{0,\pi}+B&t
\end{pmatrix}\,.
\end{equation}
This gives
\begin{equation}\label{invariantny2}
\det\bar{\mathcal{H}}(\hat\Gamma_{1,2})=B^4-(f_{0,\pi}^2-t^2-\alpha_y^2)^2+2(f_{0,\pi}^2+t^2+\alpha_y^2)\Delta^2 +\Delta^4-2B^2(f_{0,\pi}^2+t^2-\alpha_y^2+\Delta^2)\,,
\end{equation}
for the expression for the topological invariant.

For $N_y=3$ one finds, again reinserting $t$,
\begin{equation}\label{ny3h}
\bar{\mathcal{H}}(\hat\Gamma_{1,2})=
\begin{pmatrix}
0&-t&f_{0,\pi}+B&0&-\im\alpha_y&\Delta\\
-t&f_{0,\pi}+B&-t&\im\alpha_y&\Delta&-\im\alpha_y\\
f_{0,\pi}+B&-t&0&\Delta&\im\alpha_y&0\\
0&-\im\alpha_y&\Delta&0&t&-f_{0,\pi}+B\\
\im\alpha_y&\Delta&-\im\alpha_y&t&-f_{0,\pi}+B&t\\
\Delta&\im\alpha_y&0&-f_{0,\pi}+B&t&0
\end{pmatrix}\,,
\end{equation}
and hence
\begin{eqnarray}\label{invariantny3}
\det\bar{\mathcal{H}}(\hat\Gamma_{1,2})&=&B^6-B^4(3f_{0,\pi}^2+4t^2-4\alpha_y^2+3\Delta^2)
-(f_{0,\pi}+\Delta^2)\left(f_{0,\pi}^4+2f_{0,\pi}^2(\Delta^2-2t^2-2\alpha^2)+(\Delta^2+2t^2+2\alpha_y^2)^2\right)\nonumber\\&&
+B^2\left(3f_{0,\pi}^4+4(t^2-\alpha_y^2)^2+8t^2\Delta^2+3\Delta^4+f_{0,\pi}^2(6\Delta^2-8\alpha_y^2)\right)\,,
\end{eqnarray}
\end{widetext}
for the expression for the topological invariant.

\section{No transverse SO: $\alpha_y=0$}\label{app_array}

Clearly for weak transverse $\alpha$ the phase diagrams can be understood as a perturbation from the case of wires which are transversally coupled only by hopping, without spin orbit interaction along the transversal direction\cite{Tewari2012}. Here we provide some details corresponding to arrays of wires with $\alpha_y=0$, using the same notations as in the main text.
An array of wires coupled only by a hopping term has been studied before. Here we simply rederive the topological invariant using the same method as in Sec.~\ref{sec_topinv}. After a Fourier transform with periodic boundary conditions (PBCs) imposed along $x$ and OBCs along $y$, the system can be written as $H=\sum_{\vec k}\Psi^\dagger_{\vec k}\mathcal{H}(\vec k)\Psi_{\vec k}$ with
\begin{equation}
\mathcal{H}(\vec k)=\begin{pmatrix}
f(\vec k)+B & \CL_{\vec k} & -\Delta & 0\\
 \CL_{\vec k}^*& f(\vec k)-B & 0 & -\Delta\\
 -\Delta  &0 & B-f(\vec k) & \CL_{-\vec k}^* \\
0&  -\Delta &  \CL_{-\vec k}& -f(\vec k)-B
\end{pmatrix}\,,
\end{equation}
where
\begin{equation}
f(\vec k)=-t(\cos[k_x]+\cos[k_y])-\mu\,,
\end{equation}
and
\begin{equation}
\CL_{\vec k}=-\im\alpha\sin[k_x]
\end{equation}
is the spin-orbit term which vanishes at the two time-reversal-invariant (TRI) points $\Gamma_{(1,2)}=0,\pi$. Here $k_x=2\pi n/N_x$ are the bulk momenta with $n=1,2,\ldots N_x$ and $k_y=\pi m/(N_y+1)$ with $m=1,2,\ldots N_y$ are quantum numbers labeling a transverse degrees of freedom.

The Hamiltonian for any $k_y$ at the TRI momenta can be written in block diagonal form as
\begin{equation}
\mathcal{H}_{k_y}(\Gamma_i)=\begin{pmatrix}
\bar{\mathcal{H}}_{k_y}(\Gamma_i)&0\\
0&-\bar{\mathcal{H}}_{k_y}(\Gamma_i)
\end{pmatrix}\,,
\end{equation}
where in this case
\begin{equation}
\bar{\mathcal{H}}_{k_y}(\Gamma_i)=\begin{pmatrix}
B+f(\Gamma_i,k_y) & -\Delta\\
 -\Delta&B-f(\Gamma_i,k_y)
\end{pmatrix}\,.
\end{equation}
The topological invariant for each $k_y$ is then, provided $\alpha\neq0$,
\begin{eqnarray}\label{rashba_ti}
\delta_{k_y}&=&\sgn\left[\det\bar{\mathcal{H}}(\Gamma_1,k_y)\det\bar{\mathcal{H}}(\Gamma_2,k_y)\right]\nonumber\\
&=&\sgn\left\{\left[B^2-\Delta^2-f^2(0,k_y)\right]\left[B^2-\Delta^2-f^2(\pi,k_y)\right]\right\}\,.\nonumber\\&&
\end{eqnarray}
When $\delta=-1$ the system is topologically non-trivial and hosts Majorana states on the edges, for $\delta=1$ it is topologically trivial.

\begin{figure}
\includegraphics*[width=0.95\columnwidth]{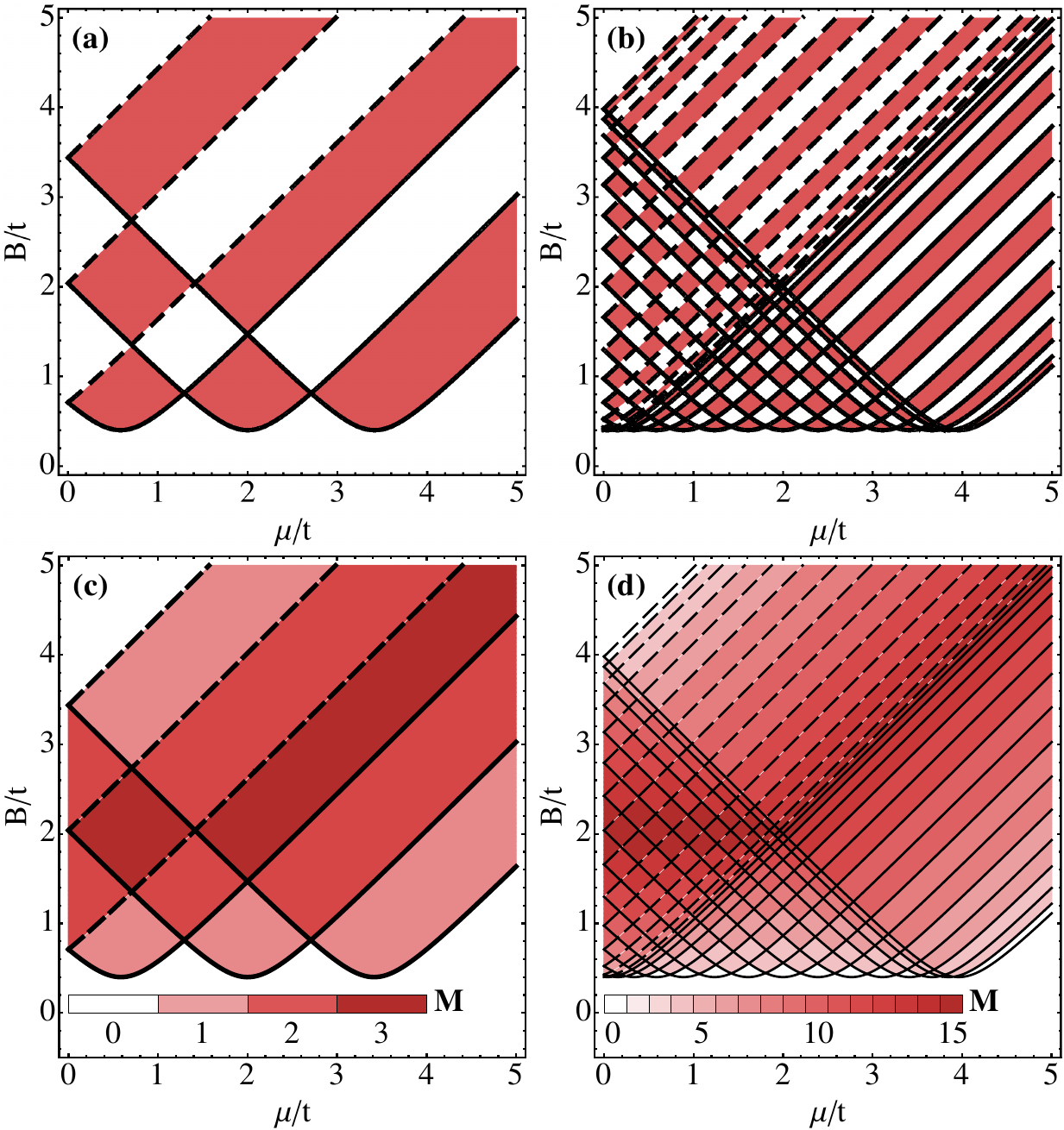}
\caption{(Color online) Topological phase diagram as a function of $B$ and $\mu$ for  (a,c) $N_y=3$ and (b,d) $N_y=15$, with transverse $\alpha_y=0$ and $\Delta=0.4$. For (a) and (b) light red is topologically non-trivial and white is topologically trivial, where we have plotted the parity of the $\mathbb{Z}$ invariant which is a $\mathbb{Z}_2$ invariant. Panels (c,d) show the number of MBS pairs, $M$.}
\label{figure8}
\end{figure}

\begin{figure}
\includegraphics*[width=0.95\columnwidth]{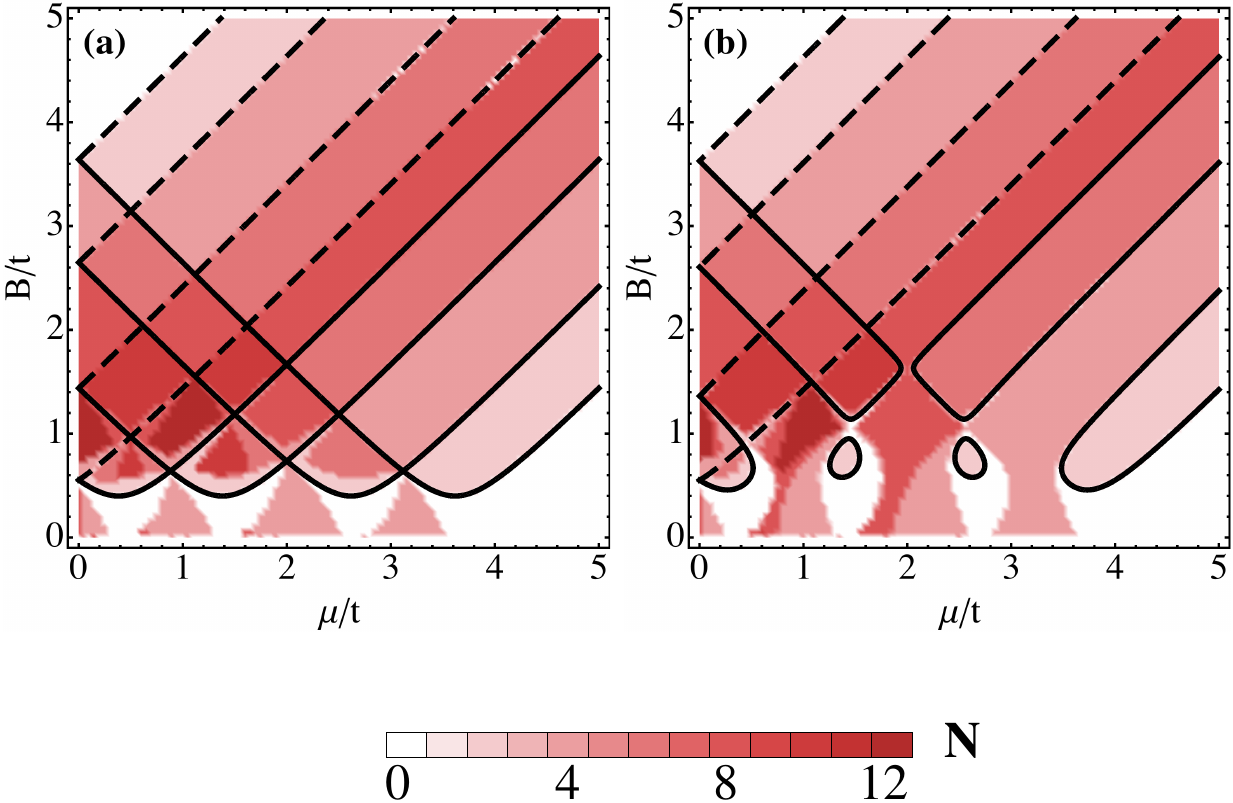}
\caption{(Color online) Number of edge state pairs, $N$, as a function of $B$ and $\mu$ for a system with $N_y=4$, $\Delta=0.4$ and (a) $\alpha_y=0$, (b) $\alpha_y=0.3$. For $\alpha_y=0$ many of these states are the protected MBS, which become gapped for $\alpha_y\neq0$.
}
\label{figure9}
\end{figure}

Fig.~\ref{figure8} shows the topological phase diagram and the Majorana number as a function of $B$ and $\mu$ for several systems with $\alpha_y=0$. Some regions of the phase diagram support a large number of MBS protected by chiral symmetry.\cite{Wang2014,Sedlmayr2015a}
In Fig.~\ref{figure9} we plot the number of edge states. Although the general pattern follows the form derived from the topological invariant, the quasi-1D wires show also a variety of unprotected edge states.

\section{Asymmetric spin-orbit coupling: weak and strong coupling limits}\label{app_asymm}

It is also straightforward to calculate the topological phase diagrams for the situation where the hopping and spin-orbit coupling is not of the same strength in the transverse and longitudinal directions. Consider Eq.~\eqref{hamiltonian} where we now allow $t_y\neq t_x$ and $\alpha_y\neq \alpha_x$. We will keep the relative strengths of the spin orbit coupling in the two directions the same, $\alpha_y/\alpha_x=t_y/t_x$, giving us just one more parameter to change: $t_y$. We can calculate the topological invariant using the same methods as in Sec.~\ref{sec_topinv} and will focus on two limits, weak transverse coupling $t_y\ll t_x$, and strong transverse coupling $t_y\gg t_x$.

\begin{figure}
\includegraphics*[width=0.95\columnwidth]{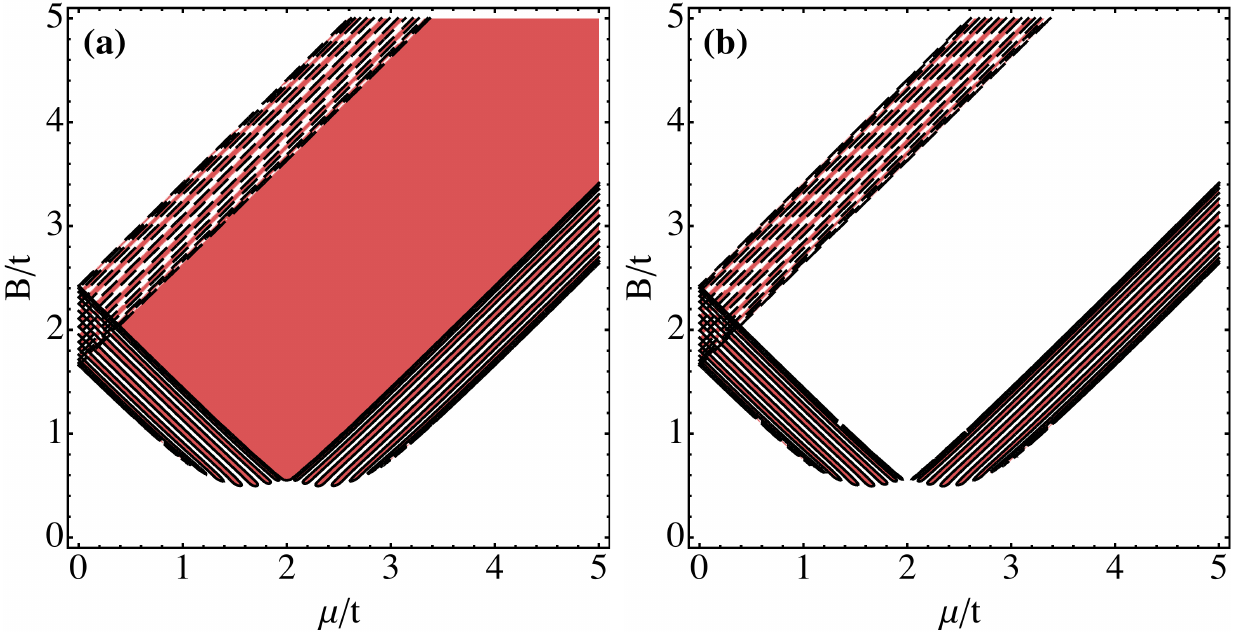}
\caption{(Color online) Topological phase diagram as a function of $B$ and $\mu$ for systems with $t_x=t$, $t_y=0.2t$, $\alpha_{x,y}=0.5t_{x,y}$ and $\Delta=0.4t$. Light red is topologically non-trivial and white is topologically trivial. Plotted are phase diagrams for (a) 15 wires, and (b) 16 wires.}
\label{figure10}
\end{figure}

\begin{figure}
\includegraphics*[width=0.95\columnwidth]{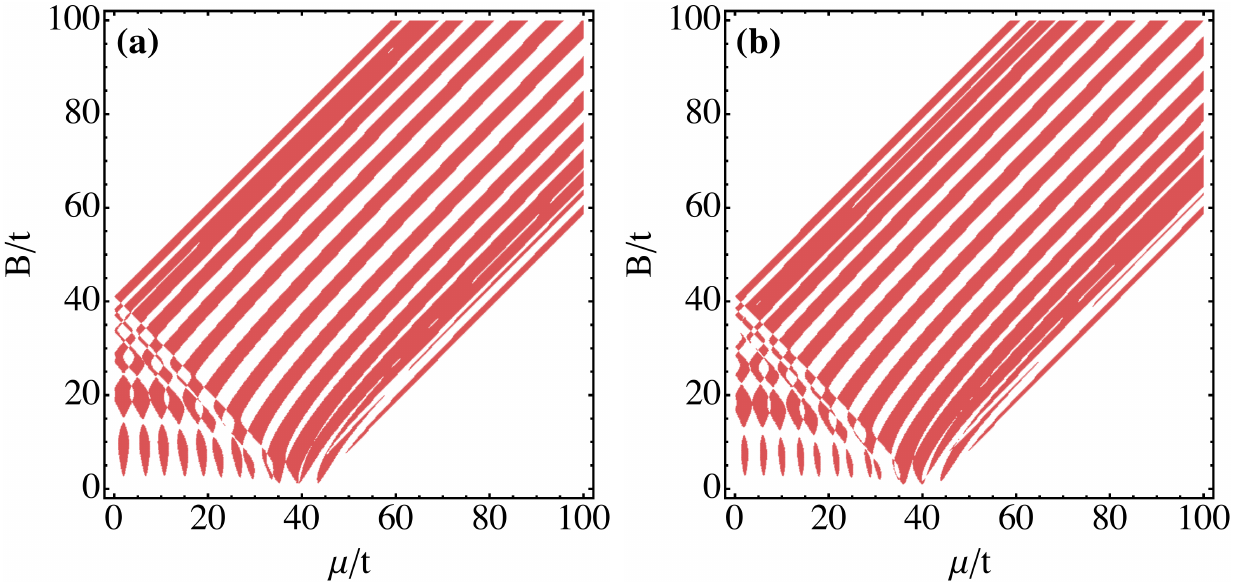}
\caption{(Color online) Topological phase diagram as a function of $B$ and $\mu$ for systems with $t_x=t$, $t_y=20t$, $\alpha_{x,y}=0.5t_{x,y}$ and $\Delta=0.4t$. Light red is topologically non-trivial and white is topologically trivial. Plotted are phase diagrams for (a) 15 wires, and (b) 16 wires.}
\label{figure11}
\end{figure}

First let us condor the weak coupling limit $t_y\ll t_x$. For an odd value of $N_y$ it is clearly seen from Fig.~\ref{figure10}(a) that the 1D phase diagram is approached as $t_y\to0$. In this limit the wires become almost decoupled in real space, each with a very similar expression for the bulk invariant. Thus in the region where all wires are topologically non-trivial there is an odd number of MBS at one end of the wire. When these states become coupled and hybridize one Majorana state must always survive. For an even number of wires this of course results in a topologically trivial region see Fig.~\ref{figure11}(b), as the even number of MBS can all combine and destroy each other.

Fig.~\ref{figure11}(a,b) show the opposite limit where $t_y\gg t_x$. In this case the wires are all strongly coupled and the situation is similar to the isotropic case $t_y=t_x$. For a large $t_y$ the principle effect is to introduce a large energy scale such that the structure in the phase diagram is of the order of $t_y$, rather than $t$. Superficially the results are similar to Fig.~\ref{figure2}.

\section{From wires to cylinders}\label{app_tubes}

\begin{figure}
\includegraphics*[width=0.99\columnwidth]{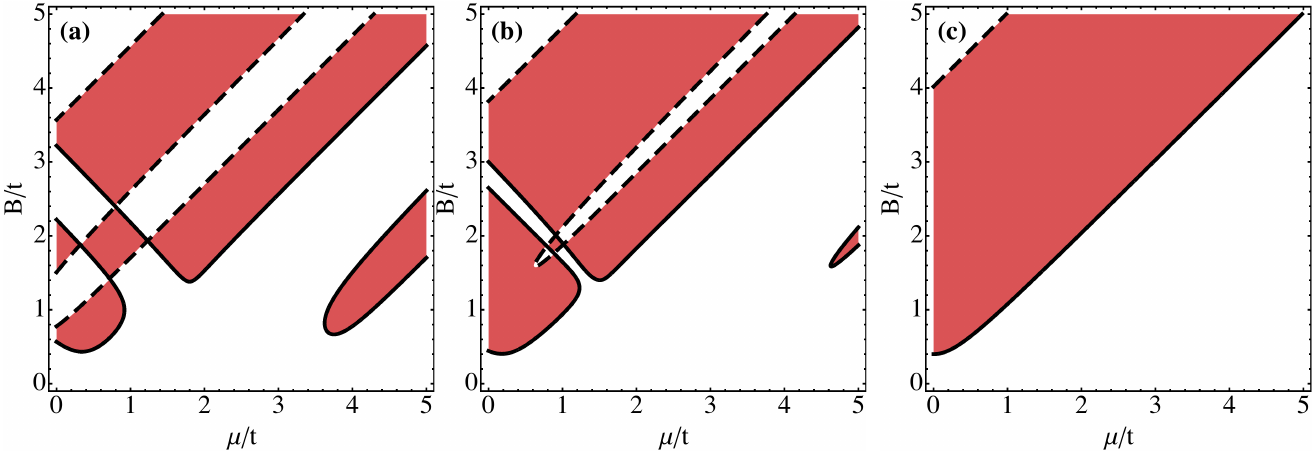}
\caption{(Color online) Topological phase diagram as a function of $B$ and $\mu$ for a system with $3$ wires with $\alpha_x=\alpha_y=0.5$ and $\Delta=0.4$. Light red is topologically non-trivial and white is topologically trivial.  Plotted are phase diagrams for (a) $t'=0.3$, (b) $t'=0.7$, and (c) $t'=1$ (the fully periodic nanotube). Note these are all 1D topological phase diagrams.}
\label{figure12}
\end{figure}

\begin{figure}
\includegraphics*[width=0.99\columnwidth]{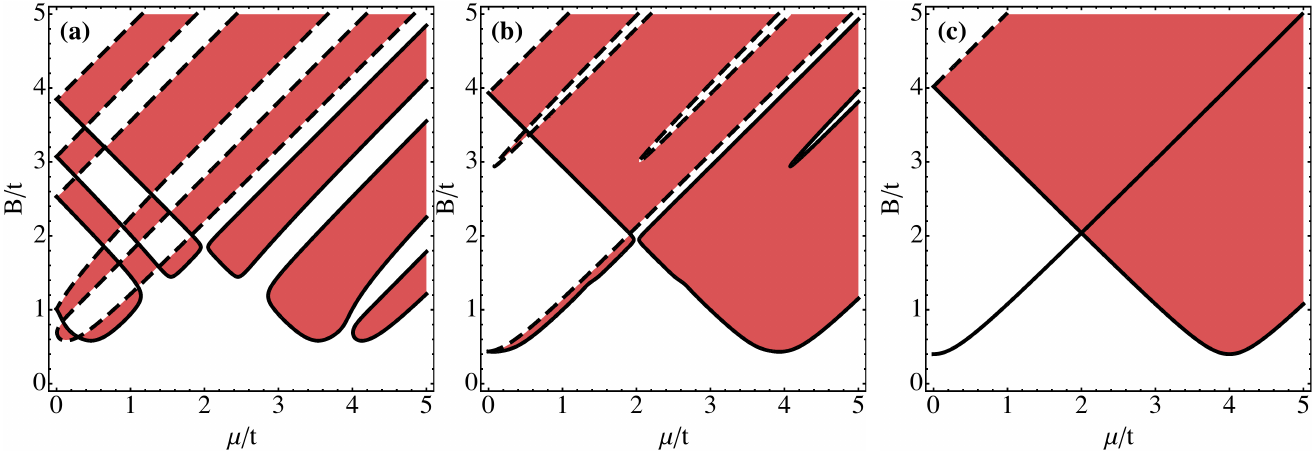}
\caption{(Color online) Topological phase diagram as a function of $B$ and $\mu$ for a system with $6$ wires with $\alpha_x=\alpha_y=0.5$ and $\Delta=0.4$. Light red is topologically non-trivial and white is topologically trivial. The coupling $t'$ has the same values as in Fig.~\ref{figure12}.}
\label{figure13}
\end{figure}

The calculation in Sec.~\ref{sec_topinv} remains valid if we consider attaching the outer wires $1$ and $N_y$ to form a periodic loop in the $y$ direction creating a cylindrical geometry. We add to the Hamiltonian Eq.~\eqref{hamiltonian} a term
\begin{equation}
H'=\sum_{x=1}^{N_x}\Psi^\dagger_{(x,N_y)}t'\left(-1+\im\alpha_y/t_y{\bm\sigma}^x\right){\bm \tau}^z\Psi_{(x,1)}+\textrm{H.c.}\,,
\end{equation}
We can then calculate the 1D topological invariant for ladders, cylinders, and for anything in between by varying the strength of this connection, see Figs.~\ref{figure12} and \ref{figure13} for two examples. For a nanotube with an even $N_y$ and $t'=1$ the phase diagram is clearly superficially equivalent to the 2D phase diagram. Nonetheless the nanotube and square lattice are defined by different topological invariants. As a function of $t'$ the phase diagram is modified by varying and merging gap closing lines at the TRI momenta. We can note that lines belonging to the two different TRI momenta can never merge as the longitudinal momentum is always a good quantum number. For $t'=1$ as $N_y\to\infty$ the system resembles increasingly accurately a 2D system with 2 boundaries. The formation of the 2D chiral Majoranas from the quasi 1D edge states, as $N_y$ is increased, has been thoroughly explored in Ref.~\onlinecite{Alicea2012}. In the case under consideration here, as $N_y$ is always finite, we remain always in the quasi-1D scenario and the edge states are modified by the coupling $t'$ without forming a band.

For an odd number of wires the nanotube possess only half the region of bulk non-trivial topology. This can be understood because for a periodic nanotube the topological invariant can be written in term of just four points corresponding to the TRI momenta for a 2D periodic square lattice. However for odd $N_y$ only two of these exist in the actual system.

\begin{figure}
\includegraphics*[width=0.95\columnwidth]{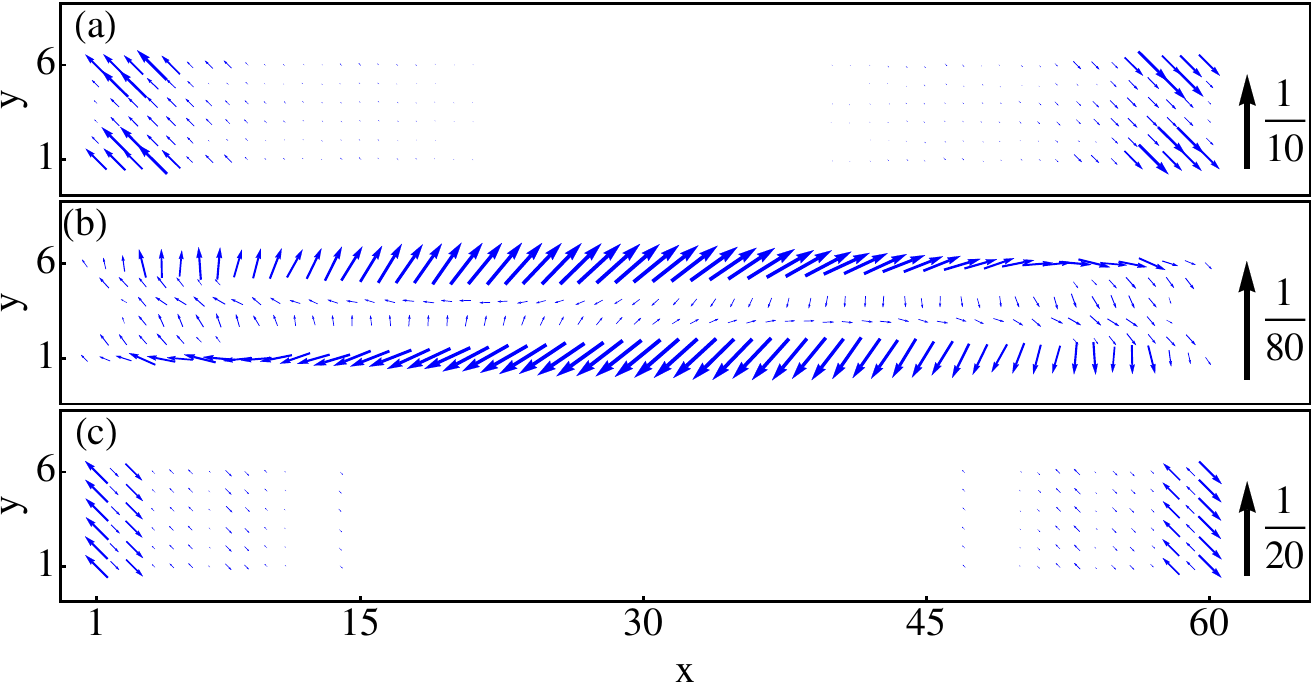}
\caption{(Color online) MP for a system with $6$ wires with $\alpha_x=\alpha_y=0.5$ and $\Delta=0.4$. We show the transition from the MBS in an open system to a non-MBS in the periodic system. The state is destroyed by leaking along the weakly coupled edges. The coupling is $t'=0,0.4t_y,t_y$ for (a,b,c) respectively. The system size is $N_x=60$ and $N_y=6$, see Fig.~\ref{figure13}.}
\label{figure14}
\end{figure}

By varying $t'$ we can switch from a quasi-1D ribbon to a quasi-1D cylinder and track the behavior of the lowest energy state using the MP. We have three interesting choices. Firstly if there is no topological phase transition and we start with a MBS then this state simply becomes translationally invariant along the transverse direction. No other change to the MP occurs. If, however we start with a MBS at $t'=0$ and pass through a phase transition on the way to $t'=t_y$ then the MBS is destroyed, see Fig.~\ref{figure14}. The lowest energy state gains a rotation in the MP along the transverse edge. Not that there are band-crossings as a function of $t'$ and hence the lowest energy state in Fig.~\ref{figure14}(c) is not continuously modifiable into Fig.~\ref{figure14}(a) by tuning $t'$. The MBS in Fig.~\ref{figure14}(a) ends up as a bulk state for $t'=t_y$ and the lowest energy state at $t'=t_y$ originates in an edge state at $t'=0$. Finally we can start with no MBS at $t'=0$ and pass through a phase transition on the way to $t'=t_y$ in which case a MBS is formed, this is simply the reverse of the previous case.


%

\end{document}